\documentclass[aip,apl,reprint,twocolumn]{revtex4-1}
\usepackage{graphicx}
\usepackage{dcolumn}
\usepackage{bm}
\usepackage{amssymb,amsmath}
\usepackage{gensymb}
\usepackage{balance}
\usepackage{textcomp}
\usepackage{times,mathptmx}
\usepackage{graphicx}
\usepackage{lastpage}
\bibstyle{apsrev}
\usepackage{hyperref}
\usepackage{color}
\usepackage[normalem]{ulem}

\begin{document}
\preprint{PREPRINT}

\title[A platform for nanomagnetism - assembled
ferromagnetic and antiferromagnetic dipolar tubes]{A platform for nanomagnetism - assembled
ferromagnetic and antiferromagnetic dipolar tubes}


\author{Igor Stankovi\'{c}}
\affiliation{
Scientific Computing Laboratory, Center for the Study of Complex Systems, Institute of Physics Belgrade, University of Belgrade, 11080 Belgrade, Serbia.
}
\email{igor.stankovic@ipb.ac.rs}
\author{Miljan Da\v{s}i\'{c}}
\affiliation{
Scientific Computing Laboratory, Center for the Study of Complex Systems, Institute of Physics Belgrade, University of Belgrade, 11080 Belgrade, Serbia.
}

\author{Jorge A. Ot\'alora}
\affiliation{Institute for Metallic Materials at the Leibniz Institute for Solid State and Materials Research, IFW, 01069 Dresden, Germany.}
\author{Carlos Garc\'ia}
\affiliation{Departamento de F\'isica \& Centro Cient\'ifico Tecnol\'ogico de Valpara\'iso-CCTVal, Universidad T\'ecnica Federico Santa Mar\'ia, Av. Espa\~na 1680, Casilla 110-V, Valpara\'iso,~Chile.}

\date{\today}

\begin{abstract}
We report an interesting case where magnetic phenomena can transcend mesoscopic scales. Our system consists of tubes created by assembly of dipolar spheres. The cylindrical topology results in the breakup of degeneracy observed in the planar square and triangular packings. As far as the ground state is concerned, tubes switch from circular to axial magnetization with increasing tube length. All magnetostatic properties found in magnetic nanotubes, in which the dipolar interaction is comparable or dominate over the exchange interaction, are reproduced by the dipolar tubes including an intermediary helically magnetized state. Besides, we discuss antiferromagnetic phases and an interesting intermediary vortex state resulting from the square arrangement of the dipolar spheres.
\end{abstract}

\maketitle

\section{Introduction}

Whether a system behaves as classical or quantum is usually determined by the ratio between its spatial dimension and the quantum-coherence-length.
Even so, there are cases where the actual size dependent behavior seems to be an illusion and transcending the scales is possible, thus allowing the study of fundamental aspects hardly accessible in the original size. Spin-ice frustration\cite{Ramirez99, Gingras11, Diep13frustrated,loehr2016defect} is an example, wherein the micro- and mesoscopic rules that govern the spin orientation of such systems can become very subtle and hard to understand. Nevertheless, \citeauthor{VenderbosPRL11}\cite{VenderbosPRL11} and \citeauthor{MelladoPRL12}\cite{MelladoPRL12} have shown that similar frustrated states can also arise in arrangements of macroscopic dipolar rotors via classical magnetic interactions, furthermore, showing phenomena that are not visible in their microscopic counterpart. In this paper, we present similar scenario of scale transcendence, relating magnetic nanotubes (MNTs) with self-assembled dipolar magnetic spheres arranged in tubular structures, named here {\it dipolar tubes}. The spheres can have radii from $10$nm to macroscopic neodymium balls. A peculiar feature of this comparison is that the tubular geometry of dipolar tubes breaks-up continuous degeneracy of the ground states in the two dimensional (2D) lattices of dipolar spheres.~\cite{PhysRevB.55.15108, PhysRevB.42.6574} As a result, we expect a number of new stable states in the tubular geometry. The curvature- induced feature opens the inquiry on its impact on the energy barriers that separate and stabilize the novel magnetic states, that will be addressed in this manuscript. 

Given the lack of exchange interaction in dipolar tubes, it is the most reasonable to compare them with dipolar interaction dominated MNTs where the exchange interaction is negligible. Since in MNTs, the dipolar dominated state is {\it circular} (magnetization polarized azimuthally), it can be expected that transcendence exists only with a similar circular state in dipolar tubes. As we see along the manuscript, the scale transcendence strikingly goes beyond this trivialization. In a ground state, the stray field created by MNTs should be minimal. This condition stems from micromagnetic considerations~\cite{weiss,hubert1969stray,aharoni2000introduction} wherein the magnetostatic energy is minimized due to dipolar part of the energy. Apart of the circular state, {\it axial} and {\it helical} ground states in MNTs have been predicted theoretically~\cite{EscrigJMMM07, LanderosPRB09}\footnote{Elsewhere in the literature axial state is also referred to as mixed and helical as transition state, cf. Ref.~\cite{LanderosPRB09}. In this work, we chose to name the states based on magnetization in the center of the tube.} and confirmed experimentally~\cite{WeberNL12,RufferNANOSCALE12,BuchterPRL13,BuchterPRB15,WyssARXIV17}. In the axial state, the magnetization is parallel to the nanotube's axis in the center of the tube and gradually turns into circular magnetization at the nanotube ends. In the helical state, the magnetization is never completely aligned with the tube's axis resulting in a circulating component of the magnetization.~\cite{EscrigJMMM07, LanderosPRB09}. The helical and axial states spontaneously emerge when the dipolar interaction is comparable to exchange interaction. 
Both states occur when nanotube's radius is a few tens of times larger than the exchange length. The axial state appears when the MNT length is around two to three orders of magnitude larger than its diameter, whereas the helical state is a transition configuration to circular state that appears when the length of the tube is further reduced. Most of the previous works on magnetic nanotubes have focused on magnetic configurations as function of geometry for specific material parameters. Only recently \citeauthor{Salinas2018}~\cite{Salinas2018} applied a generic model to discover that helical phases posses a high level of the metastability relevant for magnetization reversal modes. Still, the origin of small energy differences between the states remained unclear. The axial and helical states create small stray~\cite{LanderosPRB09} or exponentially decaying in case of infinite structures, and therefore they could also exist in dipolar tubes~\cite{StankovicSM2016}. 

The cylindrical magnetic geometry of MNTs has advantages for applications despite evident fabrication problems. In fact, the elongated geometry, azimuthal symmetry, and curvature of nanotubes bring reproducibility, robustness, and extra stability to nanotube's equilibrium states and magnetization dynamics~\cite{StreubelJPDAP16, YanAPL11, ChenJMMM07, ChangPRB94, EscrigJMMM07, LanderosPRB09, OtaloraPRL16}, which makes MNTs attractive for buffering, transport and processing information using their equilibrium states, domain wall dynamic and spin-waves excitations. In this sense, the proper understanding and characterization of equilibrium states in MNTs is, thus, a mandatory task. Under this scenario, mimicking the magnetic equilibrium features of nanotubes with dipolar tubes can facilitate and encourage developments towards alternative techniques intended to reduce the complexity of experiments. The minimal energy structures of dipolar particles have been investigated in recent theoretical studies~\cite{Messina_PhysRevE_2014, SelfAssem2016}. The tubular form of ground state together with outstanding self-assembly properties of dipolar particles~\cite{Morphew1,Morphew2} present motivation for their application as a platform for testing concepts with MNT. For instance in experiments, tubular and helical architectures with dipolar particles were obtained via DNA ligations~\cite{origami1, DNAmediatedZhang2013}, confinement~\cite{LipidNanotubescm701999m,nanotubeCNTjp0544890,nanotubeCNTnn5040923}, bulk interactions - magnetic Janus colloids~\cite{MagneticJanusYan2012}, and asymmetric colloidal magnetic dumbbells~\cite{Zerrouki_Nature_2008}. Another interesting system with respect to magnetic order are two dimensional self-assembled super lattices of magnetic cubes. The magnetic cubes are synthesized with two most probable orientations: axial [001]~\cite{chen2006synthesis} and along principal diagonal of the crystal, i.e., cube,
[111]~\cite{huaman2011novel}, but the possibility of a less trivial
orientations should not be discounted. As result of interplay of square packing and magnetization defined by their crystal structure,  axially magnetized anti-ferromagnetic states in case of [001] and {\it vortex} in case of [111] magnetized cubes~\cite{Singh_Science_2014, MehdizadehTaheri14484}. At this point, we would like to draw attention to two recently developed techniques with which dipolar tubes could be realised: ($i$) Two-photon lithography~\cite{twophoton1} nano-printers can fabricate complex three-dimensional structures with the resolution of up to 300nm. The technique was used to fabricate nanostructures out of polymer, metallic~\cite{twophoton2}, and recently magnetic~\cite{twophoton3} materials. Such printed structures could be used as a template for self-assembly of magnetic particles with rhombic and square lattices. ($ii$) The second technique comes from micro-fluidics. The tubular structures of magnetic particles can be created by the conformal covering of the cylindrical conductive wire surface by assisting the self-ordering process of magnetic microspheres~\cite{kulic} via the application of a circular electromagnetic field induced by an injected electrical current along the wire. 

With the aim of addressing our results, linking self-assembly, geometry, and magnetization states in dipolar tubes, this paper is organized as follows:   Section~\ref{method}  introduces dipolar interaction model and methods used. We discuss self-organization on cylindrical confinement in Section~\ref{selforganization},  and {\it in silico} hierarchical degeneracy breakup of the infinite square and triangular lattices with an introduction of curvature in Section~\ref{magnet}. We also present a systematic study of the ground state configurations and energies resulting from the interplay between the tube's length and curvature for triangular and square arrangements in Secection~\ref{magnet}.  Final Section~\ref{conclusion} gives conclusion and outlook.

\section{Models and methods}
\label{method}
\subsection{Magnetic interaction model}
\label{dipoleSec}

Magnetic nanoparticles can have complex coupling involving both dipolar and exchange interactions. The atomic exchange interaction is relevant up to a length scale of $10$nm~\cite{LIN2006100}. Thus, dipolar coupling dominates in the formation of the structures on the length scales $10$nm-$100$\textmu m, with many potential applications~\cite{origami1,DNAmediatedZhang2013, MagneticJanusYan2012, LipidNanotubescm701999m,nanotubeCNTjp0544890,nanotubeCNTnn5040923}. We characterize the system using dipole-dipole interaction potential: it is assumed that each particle carries identical dipolar (magnetic) moment with magnitude $m=|\vec m_i|$,
where $\vec m_i=(m^x_i,m^y_i,m^z_i)$ defines the dipolar moment of particle $i$. The potential energy of interaction $U(\vec r_{ij})$ between two point-like dipoles with centers located at $\vec r_i$ and $\vec r_j$
can be written as:
\begin{equation}
  U(\vec r_{ij}) = \frac{\mu_0}{4 \pi} \displaystyle \left[ \frac{\vec
  m_i \cdot \vec m_j}{r_{ij}^3} - 3 \frac{(\vec m_i \cdot \vec r_{ij}) (\vec  m_j \cdot \vec r_{ij})}{r_{ij}^5} \right],
  \label{dipole}\end{equation}
for $r_{ij} \geq d$ or $\infty$ otherwise, where $r_{ij}=|\vec r_{ij}|=|\vec r_j - \vec r_i|$ and $d$ is particle's diameter. It is convenient to introduce the energy scale defined by
$U_{\uparrow\uparrow} \equiv \mu_0 m^2/4 \pi d^3$ that physically represents the repulsive potential value for two parallel dipoles at contact standing side by side, as clearly suggested by the notation.
Thereby, the total potential energy of interaction in a given structure $U_{\rm tot}$ is given by
  \begin{equation}
  U_{\rm tot} = \sum_{i>j}U(\vec r_{ij}).
  \end{equation}

One can then define the reduced potential energy of interaction $u$ (per particle) of $N$ magnetic spheres. It reads:
  \begin{equation}
  u_N = \frac{U_{\rm tot}}{U_{\uparrow\uparrow}N},
  \end{equation}
which will be referred to as the {\it cohesive energy}. The cohesive energy of a particle is directly related to the energy required to take it out from the structure. Lower cohesive energy means it takes more energy to disintegrate structure. The higher is the absolute value of cohesive energy the more stable is the structure. For the particular two particle head-to-tail configuration (i.e., $\rightarrow\rightarrow$), we get $u_2 = -1$ per particle.

There is significant flexibility in tuning physical and chemical properties of magnetic particles. In particular colloids can be synthesized either from pure magnetic material like hematite with a small spontaneous magnetization ($I^{\rm s}_{\rm FeO}\approx2.2$kAm$^{-1}$), or large in case of magnetite or cobalt ferrite ($I^{\rm s}_{\rm CoFe}\approx480$kAm$^{-1}$)~\cite{params_Sacanna2012,hematite}. In case of core-shell particles, a design freedom is obtained by the adjustable core to shell ratio. Here, in particular, we consider all magnetic particles have the same magnetic moment. We assume that particles are silica-hematite core-shell particles with outer diameter $d=50$\textmu m and hematite core $d_{\rm core}=10$\textmu m~\cite{fang2008monodisperse}. Assuming a single domain particle behavior, magnetic moment $m$ is expressed as $m = I_{\rm FeO}^{\rm s}v =1.15$A\textmu m$^2$, where $v\approx$500\textmu m$^3$ is the volume of the magnetic part of the particle. The referent magnetic interaction energy $U_{\uparrow\uparrow}=10^{-18}$J, i.e., the maximum magnetic attraction that spheres can generate at contact, was estimated to be $256k_{\rm B}T$, where $T=300$K is the temperature and $k_{\rm B}$ is the Boltzmann's constant. The maximal magnetic field generated by one particle at the center of mass of the other particle (placed side by side) is $B_{\rm 0}=\mu_{\rm 0}m/(2\pi d^3)=1$\textmu T. The size of the magnetic core has a strong influence on energy scale: for $d_{\rm core}=20$\textmu m, we would obtain magnetic moment $m = I_{\rm FeO}^{\rm s}v = 9.2$A\textmu m$^2$, and magnetic energy depends on square of magnetic moment $U_{\uparrow\uparrow}=67\cdot10^{-18}$J, i.e., $1.6\cdot10^4k_{\rm B}T$. As result one could tune level of degeneracy described in latter sections with the size of the core. Also, by changing core to shell ratio we tune the balance of interaction between particles and of particles with the field created by a conducting wire.

\subsection{Isotropic interaction}

When the dipolar coupling is strong, such as in nanocrystals, the particle assembly is determined unequivocally by the dipolar coupling and the particle shape. Here we are interested in moderately interacting magnetic particles since we want to avoid the spontaneous formation of the clusters. Self-assembly requires to take advantage of forces that dominate on the micron scale and below (magnetic, contact, and van der Waals), resulting in different device designs and functionalities~\cite{Min2008}. 

We describe the effect of isotropic contact and van der Waals interactions between the spherical particles using a minimal model, i.e., as soft-core beads, that interact isotropically by means of a truncated and shifted Lennard-Jones potential. The interaction is defined as: $U^{\rm cut}_{\rm LJ}(r)=U_{\rm LJ}(r)-U_{\rm LJ}(r_{\rm cut}), r<r_{\rm cut}$ and $U_{\rm cut}=0, r\geq r_{\rm cut}$, where $r_{\rm cut}$ is the distance at which the potential is truncated, and $U_{\rm LJ}(r)$ is the conventional Lennard-Jones (LJ) potential, i.e., $U_{\rm LJ}(r) = -4\epsilon[(\sigma/r)^{12}-(\sigma/r)^{6}]$. The parameter $\epsilon$ corresponds to the energy scale of the interaction and $\sigma$ is related to the characteristic diameter of the beads $d$. i.e., $\sigma=d/2^{1/6}$ and $d=50$\textmu m. The choice of the $r_{\rm cut}$ value determines nature of the potential $U^{\rm cut}_{\rm LJ}(r)$: {\it repulsive} also known as Weeks-Chandler-Andersen (WCA) potential for truncated Lennard-Jones potential in minimum $r_{\rm cut} = 2^{1/6}\sigma$, and {\it attractive} for a commonly chosen $r_{\rm cut} = 2.5\sigma$. Presence of attractive interactions between particles is reminiscent of Stockmayer fluid, a simple and convenient model for representing ferrofluids~\cite{stockmayer1,stockmayer2} or lattice of particles stabilized by dipolar coupling~\cite{DipolarCoupling_nn301155c}.

The colloidal structures analyzed here are modeled to represent the colloidal magnetic particles that have a iron oxide inclusions inside the silica shell:  {\it attractive} part of isotropic interaction, we choose a weak interaction between particles~\cite{DipolarCoupling_nn301155c}, $\epsilon_a = 3.5\cdot10^{-19}$ J ($\epsilon_a =0.3U_{\uparrow\uparrow}$ for $d_{\rm core}=10$\textmu) and $r_{\rm cut}=125$\textmu m for particles with $d=50$\textmu m diameter. The value of the sphere repulsive contact potential is taken $\epsilon_r=7\cdot10^{-16}$J for particles with the same $d=50$\textmu m diameter (i.e., $\epsilon_r=10U_{\uparrow\uparrow}$). The magnitude of the attractive part of potential and interaction range can be varied by controlling the colloidal charge number or surface composition~\cite{params_Dijkstra2005PRL}.

We study the system by means of Langevin molecular dynamics computer simulations (described in following subsection): our spheres are represented by WCA potential, and carry a magnetic point dipole in their centers. A weak isotropic van der Waals (vdW) attraction between spheres is included for a more realist approach to experimental solution of colloidal particles. In the experiment, colloidal particles are stabilized against irreversible agglomeration by vdW forces, either by polymers grafted to the surface, or manipulating the ionic content of the fluid. The vdW attraction between spheres provides additional stability to the lattice composed by assembled tubes after the electromagnetic field has been switched-off. As such, our results can be scaled to different shell materials, i.e., polystyrene and silica oxide, but the conclusions of this minimal model should be generic.

\subsection{Interaction with conductive wire}
We place the conductive wire in a suspension of spherical magnetic particles. The conductive wire is an elegant way to cover the cylindrical surface with magnetic particles. Such system has been recently implemented by ~\citeauthor{kulic} with paramagnetic particles~\cite{kulic}. When replacing paramagnetic with magnetic particles, the magnetic fields of the ferromagnetic particles and electromagnetic field generated by the conductive wire become independent. As result, we obtain an additional tuning parameter - a ratio between magnetization of the particle and electromagnetic field or current of the wire. Still, the system parameters should be carefully selected to avoid the formation of kinetically trapped clusters or arcs of particles attached to the wire. 

We consider a situation in which colloidal suspension is placed in the vicinity of the current conducting wire. A wire with outer diameter $2R_{\rm w}=$50-100\textmu m is connected along the z-direction. In order to generate a electromagnetic field able to attract particles at the surface of the wire, significant currents must go through the wire, giving rise to fields of several mT at the wire surface. Recent similar experiments with paramagnetic particles indicate that a 50\textmu m wire can support currents of up to $0.5$A for several minutes and up to $0.8$A for a short times, creating electromagnetic fields up to $B=3.2$mT~\cite{kulic}, i.e., $B_{\rm w}=\mu I/\pi(2R_{\rm w}+d)$ for $d_{\rm w}=$50\textmu m wire and $d=50$\textmu m particle.\footnote{To favor the comparison with previous research in paramagnetic particles, we present here results for wires with $2R_{\rm w}=$100 and 130\textmu m and particles with 50\textmu m diameter. The ohmic heating limits the current through wire and is proportional to the square of the current and wire radius. Since the electromagnetic field ($B$) is linearly proportional to the current and inversely proportional to the distance of centers of the particle and wire, the power is $P\propto B^2(1+d/R_{\rm w})^2$. Therefore, we should note that increase of the wire diameter allows higher electromagnetic field for similar dissipation.} We will show how the interactions between the particles and of particles with the wire can be balanced to obtain single wall tubes. Our design based on ferromagnetic particles has a freedom of tuning ratio between two magnetic forces: interparticle magnetic force $F_{\rm mm}$ and electromagnetic force between ferromagnetic particles and the conductive wire $F_{\rm mI}$, i.e.,magnitude of magnetic force between two particles depends on square of their magnetic moments and the force between particles and conducting wire depends linearly on the magnetic moment. At the same time, magnetic moment is proportional to the cube of the core's diameter allowing variation of the ratio for up to three orders of magnitude, therefore this ratio can be anywhere between $F_{\rm mI}/F_{\rm mm}=1-1000$ (the single wall tubes will be created only at the higher ratios).

\subsection{Langevin molecular dynamics}

Langevin molecular dynamics (MD)\cite{PLIMPTON19951} was used to study the self-assembly in the vicinity of the wire under the influence of an electromagnetic field of uninsulated conductive wire. The total force of implicit solvent on each particle has the form:
$\vec{f}=\vec{f}_{\rm c}+\vec{f}_{\rm f}+\vec{f}_{\rm r}$, where $f_{\rm c}$ is the conservative force of inter-particle interactions and of particles with the wire,  $f_{\rm f} = -(m/\xi) v$ is a frictional drag or viscous damping term proportional to the particle's velocity, and
$\langle f_r\rangle=\sqrt{k_{\rm B} T m / \xi}$ the random Brownian force of the solvent. The random force term is treated as a Gaussian process that adheres to the fluctuation-dissipation theorem. The rotational degrees of freedom are, of course, governed by the equations of motion for torque and angular velocity of a sphere. Since evolution in time is not of primary concern in this study, the values of mass, inertia and translational/rotation friction coefficients are physically inconsequential to the final state of the system. An estimate of time, per MD step, can be obtained for 50\textmu m-sized colloidal particles with $d_{\rm core}=20$\textmu m as
$t=\sqrt{M_{\rm sp}d^2/U_{\uparrow\uparrow}}=80$ms (mass of the core-shell hematite/silica particle $M=10$\textmu g). The total length of the MD simulation was thus estimated to be of the order of 15 minutes (i.e., about $1000$s).

\subsection{Energy minimization}

The energies of finite tubes were independently computed using $10-10^3$ initial configurations with random magnetization (depending on the size of the system). The procedure included two steps: in the first step 'overdamped' equations of rotational motion of each particle were integrated with respect to the torque excreted on a particle (same equation as in Langevine molecular dynamics equations as in the previous section). The parameters used correspond to motion in highly viscous fluid where angular velocity is proportional to torque, i.e., in the limit where no acceleration takes place, in order to avoid any oscillations. In the second step, the resulting configurations were used as the input to a rigorous conjugate gradient minimization algorithm~\cite{barrett1994templates}. The second step was required since 'overdamped' rotational motion converge slowly towards the ground state\footnote{A simple example of discrete ground state is two dipole case: put two dipoles next to each other and let them orient freely in three dimensional space, they will align their moments in a head to tail configuration (coaxially).} for prescribed geometry. The energies of resulting configurations were compared - about $10\%$ of configurations had same energy, in the limit of numerical precision of about $\delta u/u=10^{-7}$, corresponding to ground state. The minimization procedure always finds dipole moments tangential to cylindrical surface, cf. Ref.~\cite{StankovicSM2016}.

  \begin{figure}[th]
  \centering
  \includegraphics[width = 8 cm]{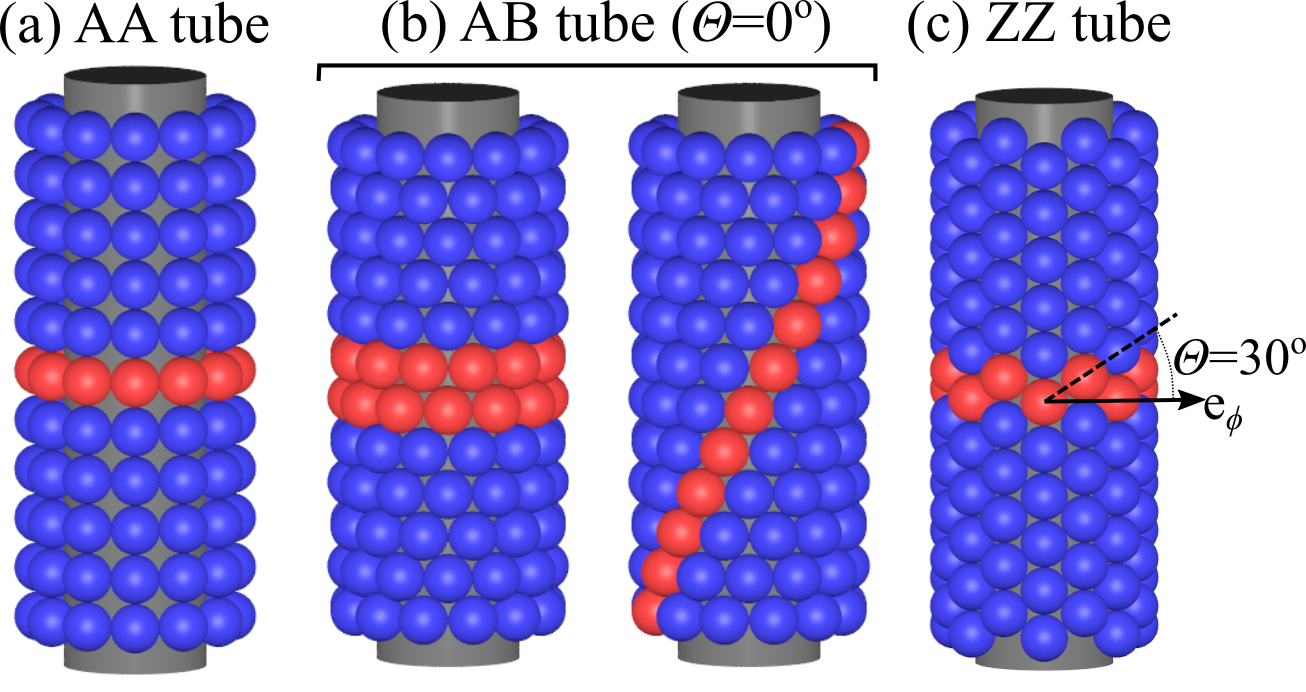}
  \caption{Illustration of (a) AA, (b) AB, and (c) ZZ tubes. The tubes are wrapped around the confinement cylinder. Tubes can be created via ring stacking (highlighted). A single ring is enough in case of AA and ZZ tubes. We show AB tubes can be created in two ways. First way is by a pair of successive rings in case of AB tube (left panel, see supplementary movie1). Second way is by wrapping of the ribbon with triangular lattice on cylindrical confinement (right panel, see supplementary movie2). In the right panel of AB tube, the edge of the ribbon with 12 threads is denoted. ZZ tube can also be created in tree ways, stacking of zik-zak  rings (pictured), wrapping 14 filaments parallel to the tube's long axis, or 14 thread ribbon oblique to the axis (see supplementary movie3 and 4). The AB tube has a chiral angle $\Theta=0^\circ$. The lattice structure of ZZ tube is triangular, like the one of the AB tube, while chirality is different, i.e., $\Theta=30^\circ$. \vspace{-1mm}}
  \label{fig:addtubes}
  \end{figure}

\subsection{Geometry of tubes}

We refer to tubes made by stacking of rings\footnote{Apart of self assembly on micro-scale, it is possible to construct tubes described in this subsection manually on the macroscopic (millimetre) scale. The  neodymium magnetic spheres are widely available and applicable for building model systems~\cite{MelladoPRL2018}. Neodymium magnets are made of a sintered alloy of iron, neodymium, and boron (Nd$_2$Fe$_{14}$B). The coercive field strength of about $10^6$A/m. Thus, the neodymium magnets can withstand much higher external magnetic fields.  The remanence of $1$ to $1.5$T is at the same time not larger than in other magnetic materials. All tube's constructed in this section can be therefore build with neodymium magnets.}. In AA-tubes all constitutive rings are exactly aligned, cf. Fig.~\ref{fig:addtubes}(a), and in AB-tubes every ring is shifted for half of the particle's diameter, in respect to its preceding ring, cf. Fig.~\ref{fig:addtubes}(b). Alternatively, AA- or AB-tubes could be generated by rolling of a square or triangular lattices on cylindrical confinement, respectively.

Particle $i-$positions in AA tubes are calculated as: $x_i = R cos (2 \pi i/N) $, $y_i = R sin (2 \pi i/N) $, and $z_i = \lfloor i/N \rfloor d$, where $\lfloor x \rfloor$ is the greatest integer function and gives the largest integer less than or equal to $x$, while $N$ is the number of particles in a constitutive ring. To simplify discussion, we refer to $N$ also as {\it curvature} since there is a correspondence with the tube's geometrical curvature $R/d=1/2\sin(\pi/N)$, e.g., we obtain $R/d=\sqrt{2}/(\sqrt{3}-1)\approx1.3$ for $N=8$ ring.

One of the ways to obtain AB tubes is stacking of a pair of two successive rings\footnote{The tubes ($n_{\rm c} = 6$) can also be created, in analogy to carbon nanotubes, by rolling a ribbon of a triangular lattice on a cylinder surface~\cite{StankovicSM2016}. The cylindrical geometry is infinite in one direction and we can, in analogy with crystal lattices, generate tubes by periodical reproduction of a curved patch (unit cell) along the helical backbone with {\it
spanning vectors} $(\vec a_1, \vec a_2)$. This curved unit cell has $n_1$ particles along $\vec a_1$ direction and $n_2$ in $\vec a_2$ direction.}. In both rings particle positions are calculated based on their index $i$: $x_i = R cos (2 \pi i/N + \theta_i)$, $y_i = R sin (2 \pi i/N + \theta_i)$, and $z_i = \lfloor i/N \rfloor \Delta z$, where $\theta_i$ is angular displacement of rings
$\theta_i=\pi \mod( \lfloor i/N \rfloor, 2 )/N$ and $\Delta z = \sqrt{d^2-2R^2 [1-\cos (\pi/N)]}$ is the displacement between successive rings along AB tube's axis and $i = {1, N_{\rm tube}}$. Total number of particles in the tube $N_{\rm tube}$ is a multiple of the number of particles in the ring $N$ and the number of rings $N_{\rm rings}$, i.e., $N_{\rm tube}=N_{\rm rings}$.

In addition to stacking of the rings, the tubes can be created by rolling a ribbon with square or triangular lattice on a cylindrical surface. Right side panel in Fig.~\ref{fig:addtubes}(b) shows an edge of the ribbon creating exactly same structure as by stacking of rings (see also supplementary movie1-4). In fact, every ordered tubular structure can be generated by reproduction of a curved unit cell along helical lines defined through curved spanning vectors in analogy to crystals in two dimensions. This curved unit cell has $n_{1}$ and $n_{2}$ particles along two spanning directions~\cite{StankovicSM2016}.

Still, there are geometrical limits for a ribbon with a defined structure (i.e., square, rhombic or triangular). Like in carbon nanotubes, ribbons of assembled particles can be rolled at specific and discrete ("chiral") angles. The chiral angle can take values $0<\Theta<30^\circ$ for triangular and $0<\Theta<45^\circ$ for square lattice, where $\Theta$ is the angle between tread of particles and tangent to cylinder radius\cite{wood2013self} (in Fig.~\ref{fig:addtubes}). Here we will demonstrate how combination of the rolling angle and radius decides the tube's properties with respect to magnetic state energies. We show AB and ZZ tubes which have different chiralities, $\Theta=0^\circ$ and 40$^\circ$ in Fig.~\ref{fig:addtubes}(b) and (c), respectively. Circular arrangement of AB tube corresponds to, so called, armchair carbon nanotube equivalent. The curvature of the two structures is also similar $R_{\rm AB}/d=1.932$ and $R_{\rm ZZ}/d=1.945$, while the number of particles in a constitutive ring is different, $N=12$ and $14$ for AB and ZZ tube, respectively. An arrangement, circular or helical in AB tube and axial or helical in ZZ tube, corresponds to possible choice of magnetization of tubes that is aligned with their lattice structure.

\section{ Amp\`{e}re force driven assembly}
\label{selforganization}

  \begin{figure}[t!]
  \centering
  \includegraphics[width = 7.5 cm]{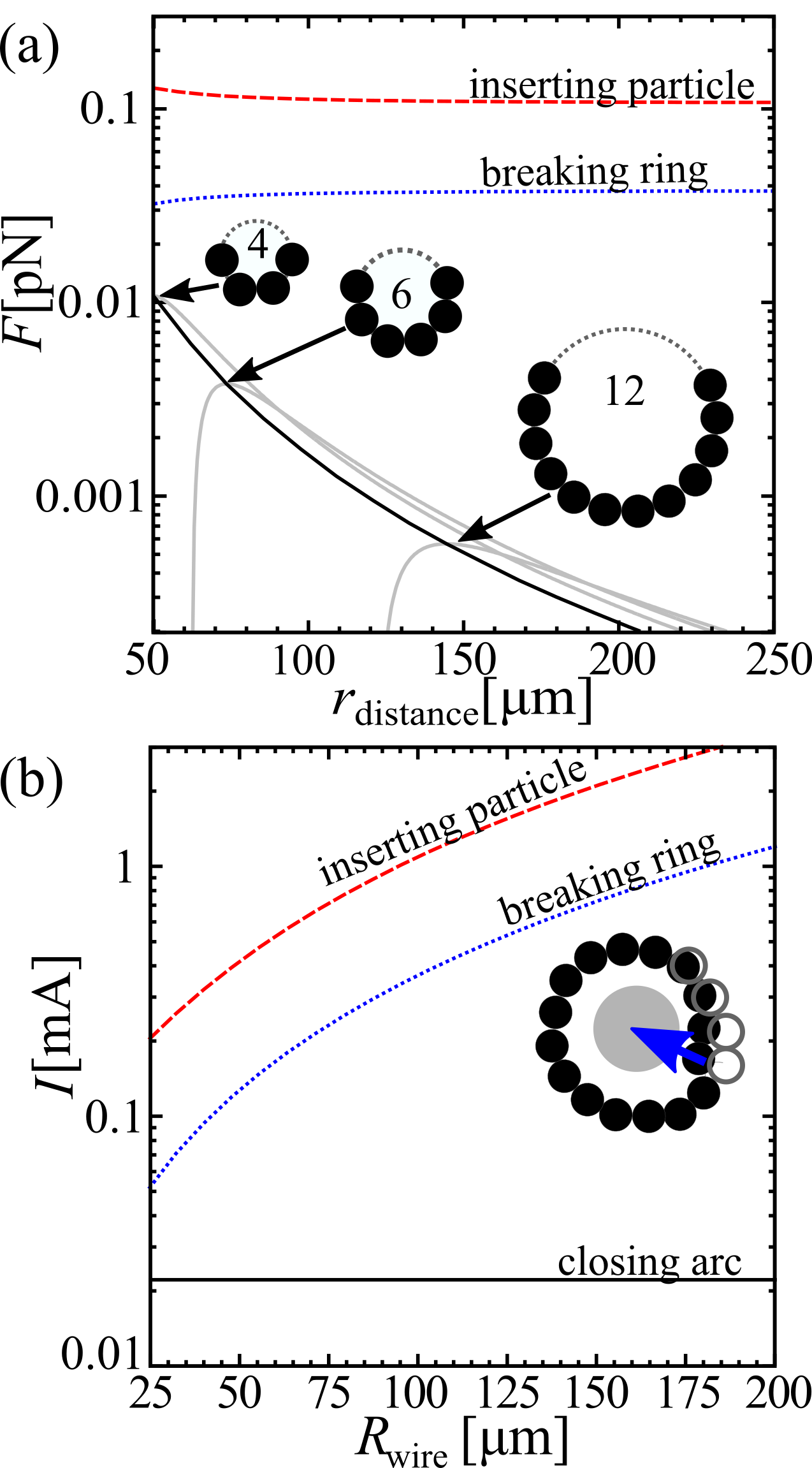}
   \caption{Critical (a) force and (b) current required to bent and close an arc of particles and form a ring, break the ring, or insert  particle between two rings.
The dependence of the forces from distance from the center of the wire $r_{\rm dist}$ is shown. The ring is broken when Amp\`{e}re's force pushes one side of ring inside (spirally deforming ring). The evolution of the force with the distance from wire $r_{\rm distance}$ is also shown in figure (a) with dotted line for $N=4, 6,$ and $12$ particles. The critical current depends on wire radius $R_{\rm wire}$ since particles become further away form the center. The magnetic moment of the particle is 1.15A\textmu m$^2$.\vspace{-1mm}}
  \label{fig:force}
  \end{figure}

The central mechanism driving adhesion of particle on a conductive wire is an interplay between dipolar forces between particles on the wire and radial attractive Amp\`{e}re force. The electromagnetic field of the conducting wire is strong enough to determine the orientation of all dipole moments. In order to obtain a single layer of magnetic particles, the Amp\`{e}re force should dominate inter-particle dipolar forces. Here we should point out that Amp\`{e}re force generated by the current in the wire depends linearly on the magnetic moment of the particles while magnetic dipolar interactions scale quadratically. Besides, there are symmetric and short-range forces between colloidal particles due to their surface design. We base our analysis on a simplest analytically tractable model for constitutive ring rearrangement and comparison with the MD simulations.

In following two sections, we first give analytic results for Amp\`{e}re force driven processes. After that, we compare these analytic results with the ones obtained by computer simulation for moderate and strong currents. We will show that only sufficiently strong current is able to pull and attach all particles to the wire's surface.

  \begin{figure*}[t!]
  \centering
  \includegraphics[width = 18 cm]{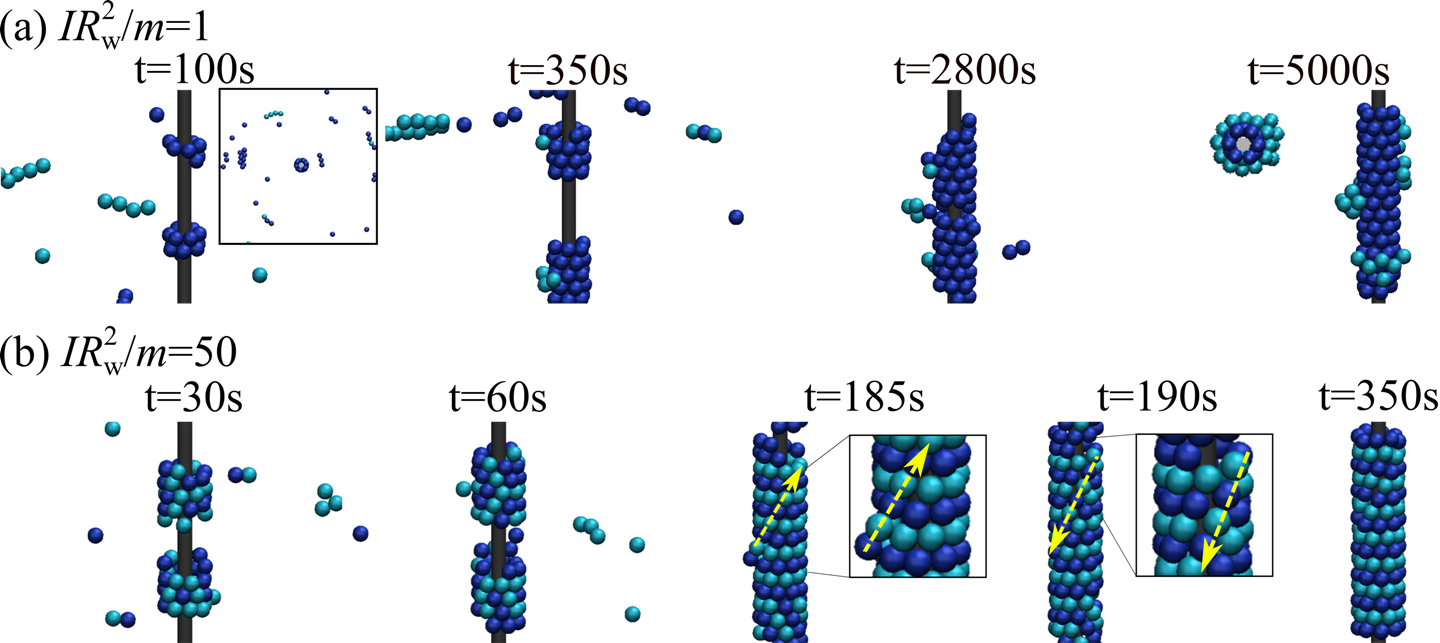}
  \caption{Snapshots of MD simulations at moderate (a) $IR^2_{\rm w}/m=1$ and strong (b) $IR^2_{\rm w}/m=50$ curents are shown. (a) For $IR^2_{\rm w}/m=1$, we observe chains form as an oriented collection of magnetic dipoles, increasingly curved by the electromagnetic field as they approach, and eventually attach to the wire. The particles in contact with wire at $t=1000$s are colored differently. The animation is given as supplementary movie5 and 6 (top and side view).  (b) At strong currents, $IR^2_{\rm w}/m=50$, Amp\`{e}re force inserts particle into the triangular lattice. Insets between t=53s and 54s show particle while entering moving (shear) tube's structure turning it into quasi stable single stranded helix. The particles belonging to successive rings are colored differently at $t=185$s to visualize this process. After some time, $t=190$s, the helix shears back into a tube with rings of magnetic particles conforming electromagnetic field lines. The animation is given as supplementary movie7. The magnetic moment of the particle is 1.15A\textmu m$^2$, the wire radius is (a) $R_w=d=50$\textmu m and (b) $R_w=1.3d=65$\textmu m, and the current is (a) $I=0.46$mA and (b) $I=23$mA. \vspace{-1mm}}
  \label{fig:formation}
  \end{figure*}

\subsection{From self-assembled chain to ring}

First agglomeration phenomenon analyzed analytically is the strength of the curved electromagnetic field needed to reduce the radius of an arc build by magnetic particles. An elongated chain (or cluster of chains) should overcome elastic barrier preventing its bending into the ring under the influence of a circular electromagnetic field. The origin of the resistance to deformation can be understood in terms of a transition from local (chain) to global energy minima, corresponding to a ring or stacking of the rings\cite{Vella_PRSA_2013,Messina_PhysRevE_2014}: For simplicity, we assume that the magnetic spheres have a magnetization that follows the curvature of the arc (i.e., part of the ring) and that the arc backbone follows electromagnetic field streamlines (i.e., co-centered with wire). 

To do so, we consider a thin wire ($r_{\rm dist}/R_{\rm wire}>1$). An arc with curvature $d/R$ can be obtained by calculating particle positions based on their index $i$ in Euclidian space: $x_i = r_{\rm dist} cos (\theta i)$, $y_i = r_{\rm dist} sin (\theta i)$, and respective magnetization $m^x_i = cos (\theta i+\pi/2)$, $m^y_i = sin (\theta i+\pi/2)$, where $\theta$ is the angular displacement of particles $\theta=2\arcsin(d/2r_{\rm dist})$ and $d$ the particle diameter. The combined resistive magnetoelastic force tries to straighten the chain and reduce its curvature $d/r_{\rm dist}$, see dashed lines for $N= 4, 6, 12$ particles in Figure~\ref{fig:force}(a). Due to the circular nature of the electromagnetic field, the curvature $d/r_{\rm dist}$ of the arc is inverse of its distance from the wire $r_{\rm dist}$. The magneto-elastic force $F$ increases up to the point when the arc ends start to attract to each other, c.f., envelope in Figure~\ref{fig:force}(a). The critical force for chain with $N=4$ particles has maximum $F=30$pN at distance $r_{\rm dist}=50$\textmu m and $N=6$ particles $F=15$pN at $r_{\rm dist}=70$\textmu m. Thereafter, the deformation of the chain becomes irreversible. The magneto-elastic force decreases with increasing curvature $d/r_{\rm dist}$ and changes the sign. The negative force means that after that point arc closes on its own. Also, one can observe that while the critical force diminishes with distance  - the highest necessary critical current is for thin wire and short arcs (i.e. for three particle arc). The critical force is inversely proportional with distance, i.e., $F\sim1/r_{\rm dist}$. The current needed to generate sufficient electromagnetic field, $I\approx0.02$A, is therefore independent on chain length.  

\subsection{Attaching particles to the surface of the wire}

We also observe that for a moderate current the long arc closes into a ring with a radius larger than the radius of the wire ($R_{\rm wire}$). How does this ring finally attach to surface of the wire? What is the critical force and current required to break the rings by spiral deformation? The transformation from a large ring to the adapted the wire diameter involves a destabilizing field able to tear apart a ring by pulling a part of it inwards to the surface of the wire ($r_{\rm dist}>R_{\rm wire}$). The energy per particle of the single ring is :
\begin{equation}
u_{\rm r}(N)=-\frac{1}{4}\sin ^3\left(\frac{\pi}{N}\right)\sum_{k=1}^{N-1}\frac{\cos \left(\frac{2 \pi  k}{N}\right)+3} {sin^3\left(\frac{\pi  k}{N}\right)}.
\end{equation}
similarly, the approximate expression for the force required to break the ring is given by (see also Figure~\ref{fig:force}(a)),
\begin{equation}
F_{\rm s}(N)=-\frac{3}{8}\sin ^3\left(\frac{\pi}{N}\right)\sum_{k=1}^{N-1}k\frac{\cos \left(\frac{2 \pi  k}{N}\right)+3} {sin^3\left(\frac{\pi  k}{N}\right)}.
\end{equation}
Since the Amp\`ere force reaches its strongest value in the wire surface, the ring will break in the vicinity at this position. We can therefore estimate critical current to be $I=2\pi(d+R_{wire})^2F/\mu_0m$, as shown in Fig.~\ref{fig:force}(b). The current required to break a ring is more than three times higher that the current needed to close an arc and rises with wire radius. Still, the increase is slow ($I\sim R_{\rm wire}^\alpha$, where $\alpha<1$) and is compensated without increase in current density through the wire. 

The magnetic particles stick (or diffuse) on top of the triangular lattice formed on the cylindrical surfaces. Following compaction, the last coming beads try to pop-in between the constitutive rings of the tube. In numerical analysis, we use the fact that a ring configuration compensates the dipole moment and the total dipole moment is zero within the ring. In far field electromagnetic field of the ring resembles a multipole, i.e., electromagnetic field drops with distance as $1/r^{N+2}$, where $N$ is the number of particles in ring. The self-screening of inter-ring dipolar interaction takes place as soon as the rings are separated more then one particle size. Therefore the change of total energy depends dominantly on the distance of the touching rings, i.e., the change of their interaction energy,
\begin{eqnarray}
u_{\rm ir}(N) & = & -\frac{1}{8} \sin ^3\left(\frac{\pi}{N}\right)
\nonumber \\
&& \hspace{-1.5cm}\times
\sum_{k=0}^{N-1}
\frac{ 2 \left\{3+\cos \left[\frac{\pi (2
k+1)}{N} \right]\right\} \sin ^2\left[\frac{\pi (2
k+1)}{2N}\right]+dz^2 S_k} {\left\{dz^2\sin^2\left[\frac{\pi (2
k+1)}{2N}\right] + \sin ^2\left(\frac{\pi}{N}\right)\right\}^{5/2}}
\end{eqnarray}
where $dz$ is the distance between touching rings and $S_k=\sum_{i=0,1,2} (-1)^i \binom{2}{i} \cos [\pi (2 (k+i)-1)/ N]$.
We estimate the force $F_{i}(N)=(6/\sqrt{2})\partial u_{\rm ir}(N)/\partial dz$ needed to push particle between two rings at contact, i.e., at distance:
\begin{equation}
dz_N^c=\sqrt{1-\frac{1}{2} \left[1-\cos \left(\frac{\pi
}{N}\right)\right] \sin ^{-2}\left(\frac{\pi }{N}\right)}.
\end{equation}
The resulting critical force and current depend on the wire radius as shown in Figure~\ref{fig:force}(a) and (b), respectively. For $R_{\rm wire}=65$\textmu m wire, on which $8$ particles of $d=50$\textmu m could form a ring, the critical current is $I=0.4$A.  

\subsection{Dynamics of assembly}

We simulate Amp\`{e}re force driven assembly of colloidal magnetic particles on cylindrical confinement. The snapshots of evolution of the configuration in time are given in Fig.~\ref{fig:formation} and animations in supplementary material as movies5-7. We model the dynamics of assembled particles with dipolar coupling in presence of the circular electromagnetic field generated by the electrical current going through a conductive cylindrical wire which at the same time serves as a geometrical constraint. 

In MD simulations at moderate currents, cf. inset in Figure~\ref{fig:formation}(a), we observe the formation of chains composed by an oriented collection of magnetic dipoles, increasingly curved by the electromagnetic field as they approach, and eventually attach to the wire. This process is schematically given in Figure~\ref{fig:force}(a). The resistance to bending increases as the particles approach the wire. The chain finally bend due to the fact that the dipoles cannot align  with both, the electromagnetic field lines and with each other's magnetic axis. In this frustrated configuration, the magnetic field of each dipole exerts a torque on all other dipoles.

At sufficiently high currents we observe that the system becomes rapidly compact, see Figure~\ref{fig:formation}(b). We also observe, between $t=185$s and $t=190$s in Figure~\ref{fig:formation}(b)\footnote{The particles are colored differently to visualize movement.}, how the late coming particle pushes the already formed triangular lattice structure forming a metastable single stranded helix. In this metastable state, helix backbone and electromagnetic field are not aligned, resulting in mechanical strain on the whole structure. At $t=350$s, we observe that system shears back into stable state (tube) with constitutive rings aligned with the electromagnetic field. 

Finally, we should note that a square lattice can only be obtained by self assembly on a square patterned surface.  This correspond also to state-of-art in literature~\cite{twophoton1}. The later one is limited to self assemble of finite sized structures. Nano-scale printing allows realization of curved conductors with a complex surface geometry and opens an interesting playground for generating different packing of magnetic spheres. The Joule heating limits the current through a 100\textmu m wire to 0.5A~\cite{kulic}. Therefore the magnetic moment of the particle is limited by total magnetic moment of the particle and should be $m<IR_{\rm w}^2/50=25A$\textmu m$^2$, taking into account whole size of the particle. Total magnetic moment of the particle is controlled by the volume of the ferromagnetic core and choice of the magnetic material. Once the dipolar tube had been formed, the particles would stay in place after the wire is removed. The tubular structures are mechanically stable also on finite temperatures (see movie8 in supplementary material). We should highlight that the wire is not only a confinement structure but an efficient way to control magnetic configuration of these tubes.

\section{Magnetization of dipolar tubes}
\label{magnet}
In this section, we analyze implication of curvature and size effects on energy landscape of triangular and square lattices. The isotropic interaction between the particles and the particles with the wire, which now serves only as rigid cylindrical confinement, does not have influence on magnetic dipole orientation.  

  \begin{figure}[th!]
  \centering
  \includegraphics[width = 8 cm]{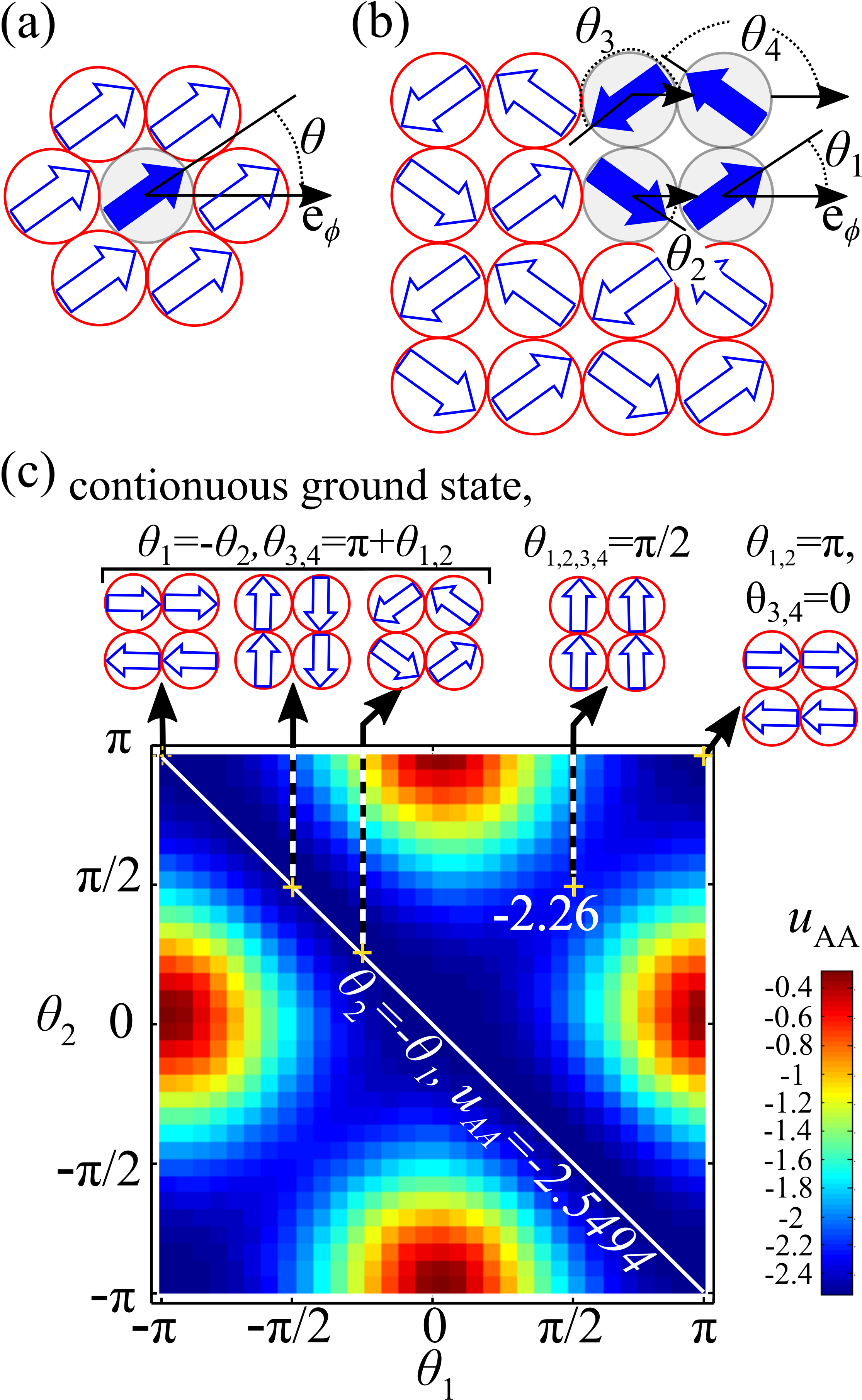}
  \caption{Visualization of degenerate states in infinite (a) triangular and (b) square lattice, i.e., respectively AB and AA packings.
  The dipoles are depicted as arrows located in the center of the spheres. In case of triangular lattice unit cell
  consists of a single particle and in case of square lattice it consists of four particles (gray). (c) A energy landscape for square lattice is shown with respect to two $\theta_1$ and $\theta_2$ out of four magnetic moments in the unit cell. Other two moments were oriented so the energy of the system is minimal. One can observe a flat valley of degenerate ground state,  $\theta_2=-\theta_1$, with energy $u_{\rm AA}\simeq-2.5494$. The saddle point which represent uniformly magnetized square plane with energy $u^{\rm sdd}_{\rm AA}=-2.26$ is also marked. The curves are drawn through the discrete points and are smooth. The results are in principle scale independent. The referent magnetic energies are $U_{\uparrow\uparrow}=10^{-18}$J and 67$\cdot10^{-18}$J, i.e., $256k_{\rm B}T$ and $1.6\cdot10^4k_{\rm B}T$, for particles with magnetic moments $m =$1.15 A\textmu m$^2$ and 9.2A\textmu m$^2$, respectively, where $T=300$K is the temperature and $k_{\rm B}$ Boltzmann's constant.\vspace{-1mm}}
  \label{fig:degen}
  \end{figure}

  \begin{figure}[th!]
  \centering
  \includegraphics[width = 9 cm]{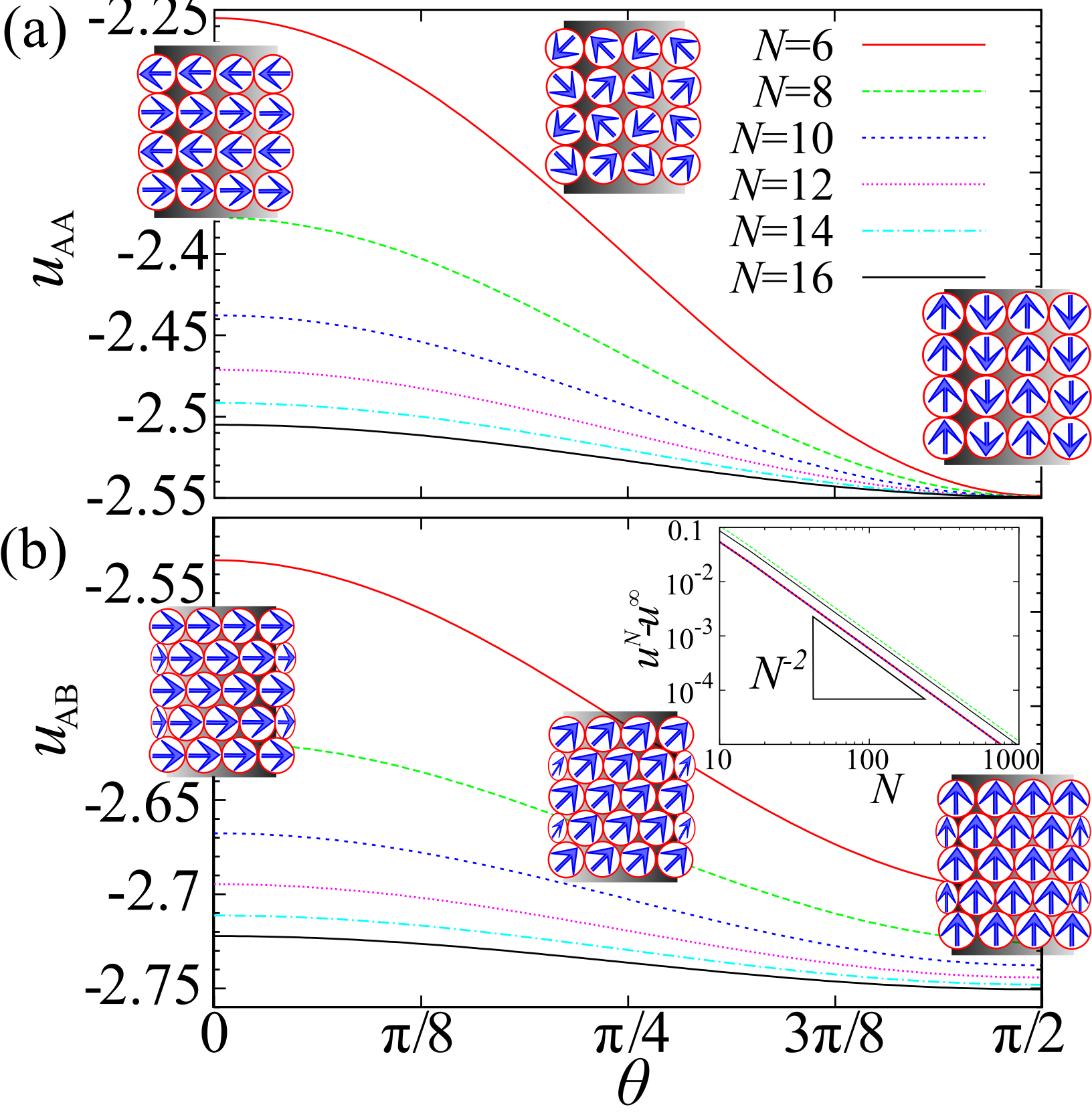}
  \caption{Dipolar cohesive energy spectrum of configurations for dipole
  orientations in Fig.~\ref{fig:degen} on a curved surface of the infinitely long tube with (a) square AA and (b) triangular AB tubes.
  Breaking of degeneracy with respect to angle $\theta$ due to the curvature, i.e., proportional to the number of particles
  in the constitutive ring $N$, is shown. The axial magnetization corresponds to $\theta=\pi/2$.
  Inset shows convergence of dipolar cohesive energies for $\theta=0$, and $\pi/4$  to infinite two dimensional plane value $u$
  (for square lattice $u_{\rm AA}=-2.5494$ and for triangular lattice $u_{\rm AB}=-2.7586$). The referent magnetic energies $U_{\uparrow\uparrow}=10^{-18}$J and 67$\cdot10^{-18}$J, i.e., $256k_{\rm B}T$ and $1.6\cdot10^4k_{\rm B}T$, for particles with magnetic moments $m =$1.15 A\textmu m$^2$ and 9.2A\textmu m$^2$, respectively, where $T=300$K is the temperature and $k_{\rm B}$ Boltzmann's constant.\vspace{-1mm}}
  \label{fig:enAng}
  \end{figure}
\subsection{Characteristics of triangular and square latices}

First, we investigate the dependence of ground state energy on magnetization. All dipoles in triangular lattice are parallel and allowed to rotate only around a fixed axis orthogonal to the plane, see Fig.~\ref{fig:degen}(a), for numerical details cf. Ref.~\cite{Brodka_MolPhys_2003}. There is a continuous ground state for any in-plane angle $\theta$ with cohesive energy value $u_{\rm AB}\simeq-2.7586$~\footnote[3]{The energy of continuous ground state independent of in-plane angle $\theta$ is $u_{\rm AB} =-2\zeta(3)+16\pi^2\sum_{k=1}^{\infty}\sum_{l=1}^{\infty}\cos(kl\pi)K_0(kl\sqrt{3}\pi)\simeq-2.7586$.}, see also Ref.~\cite{PhysRevB.55.15108,PhysRevB.42.6574}. For square two dimensional lattice, similarly, there is a continuous degeneracy of its ground state, described in Fig.~\ref{fig:degen}(b)-(c). A continuous state, in this case, involves a unit cell of four particles. The moments in a unit cell are synchronously coupled and in our notation take directions $\theta$, $\pi-\theta$, $\pi+\theta$, and $-\theta$, in anti-clockwise direction in Fig.~\ref{fig:degen}(b). The ground states found are obviously antiferromagnetic, with the total dipole moment within the cell conserved and equal to zero. The most striking is the so-called {\it vortex state} for $\theta=\pi/4$ with a fully enclosed circulation of the magnetic dipole moment within the unit cell.
The ground state cohesive energy value is $u_{\rm AA}\simeq-2.5494$~\footnote[4]{The energy of continuous ground state is
$u_{\rm AA} =-2\zeta(3)+16\pi^2\sum_{k=1}^{\infty}\sum_{l=0}^{\infty}k^2\{K_0[4k(l+1)\pi]-K_0[2k(2l+1)\pi]\}\simeq-2.5494$.}.
We will use the calculated ground state energy value as an absolute point for comparison of energies of different states in tubes with square or triangular lattice structure. We should note that both antiferromagnetic states are observed in systems of square particles as result of interplay of the magnetization defined by crystallinity of the cubes and the structure of the two dimensional super lattice. Commonly, magnetic cubes are represented by single dipoles placed in center. While this is good approximation for many systems, it only takes into account about 50\% of the total volume of the cube and is neglecting the effect of the corners. Therefore one could expect degeneracy breakup due to asymmetry of the shape of the cubes. Still, the cubes are synthesized very often with curved edges, i.e., as superballs, exhibiting a continuous transformation of shape from ideal cube to sphere~\cite{donaldson2017cube} and they are expected to self assemble in structures with square symmetry~\cite{donaldson2015directional}. An extent of degeneracy breakup remains to be analyzed during this shape-shift. 

\subsection{Degeneracy break-up with curvature}
\label{degeneracy}

Wrapping of the plane around the confinement cylinder will make the system quasi one-dimensional and break degeneracy. We will discuss repercussions of degeneracy breakup on cohesive energy for different dipole orientations. We  analyze first degeneracy breakup in infinite tubes: according to tube's cylindrical geometry, we represent dipole moment of the $i-$th particle in cylindrical coordinates like:
  \begin{equation}
  \vec{m_i} = m_{\rm i\phi} \vec{e_{\rm \phi}} + m_{\rm iz} \vec{e_z},
  \end{equation}
with constraints $m^2 = m_{\rm i\phi}^2 + m_{\rm iz}^2$ $(i=1,\ldots,N)$. The parallel component with respect to tube's axis is given by $m_z$ and the orthogonal component is $m_{\phi}$ (i.e., $m_{\phi}$ is tangential to cylinder's circumstance).
In Fig.~\ref{fig:enAng}, we follow the dependence of energy on angular parameter $\theta$, $m_{\rm iz}=m\sin(\theta)$. We find that axial magnetization (i.e., $\theta=\pi/2$) of dipole moments represents ground state for both AA- and AB-tubes, and circular magnetization (i.e., $\theta=0$) is the most unfavorable as seen in Fig.~\ref{fig:enAng}.

Between circular and axial magnetization (i.e., $0<\theta<\pi/2$), we observe a continuous increase of energy with increasing circular alignment of magnetization. These transition states, we call {\it vortex} in case of square AA tubes and {\it helical} in case of triangular AB tubes, e.g., $\theta=\pi/4$ in Figs.~\ref{fig:enAng}(a) and (b), respectively. The cohesive energy, of different configurations in Fig.~\ref{fig:enAng}, converges to continuously degenerate state with increasing curvature $N$, following power law, $u^N-u^{\infty}\sim N^{-2}$,
cf. inset in Fig.~\ref{fig:enAng}.

  \begin{figure}[th!]
  \centering
  \includegraphics[width = 9 cm]{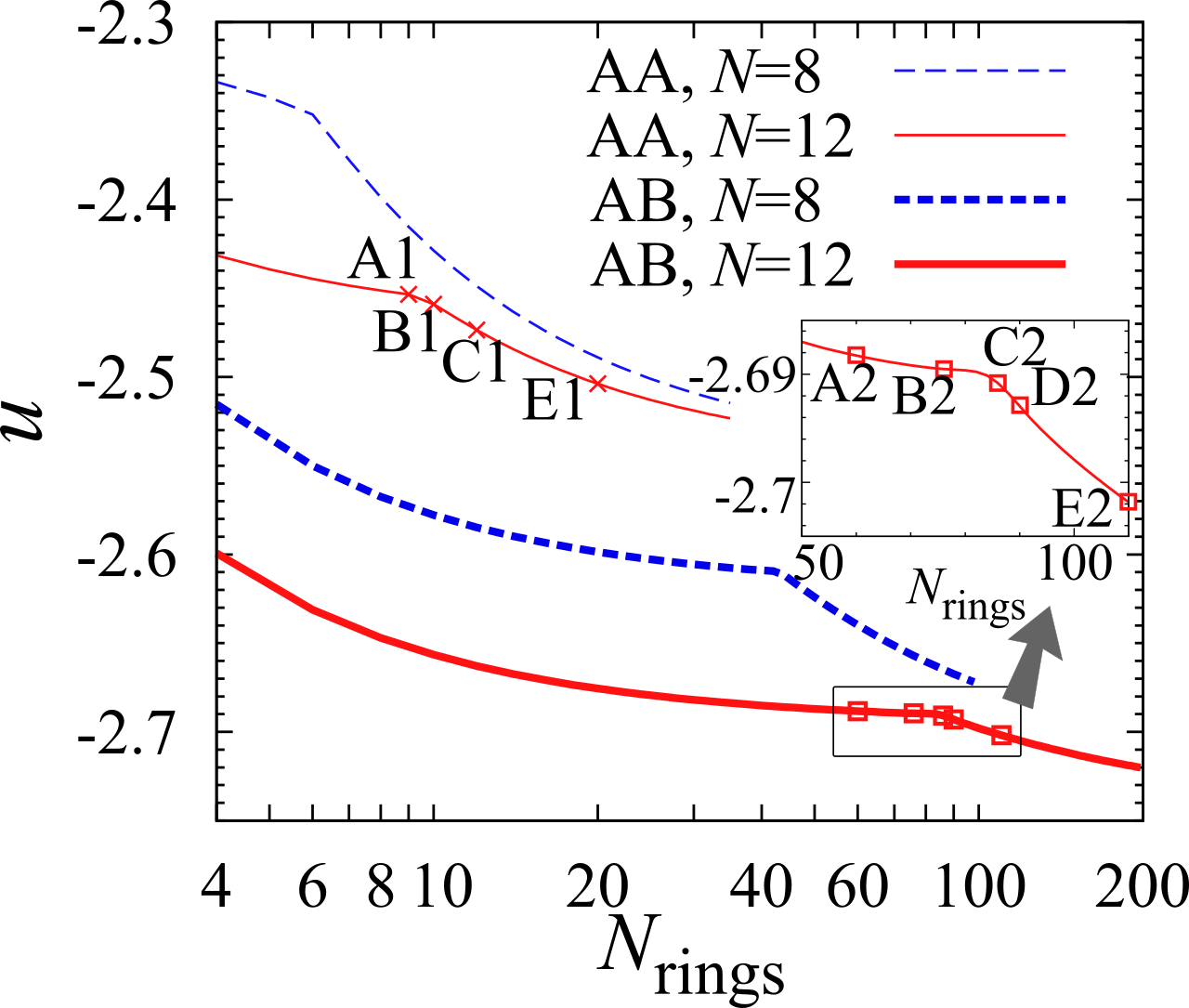}
  \caption{Reduced cohesive energy profiles $u$ as a function of the number of rings $N_{\rm rings}$
  for AA and AB tubes with curvatures $N=8$ and $12$. Configurations for points (A1), (B1), (C1), and (E1)
  are shown in Fig.~\ref{fig:AAtube} and (A2), (B2), (C2), (D2) and (E2) in Fig.~\ref{fig:ABtube}. The curves are plotted through the discrete points and serve as guide to the eye, all points lie on the curves, and only a few listed and analysed points are shown. Before points (A1) and (B2) the magnetization is ideally circular and energy decrease is only driven by addition of new rings. The results are in principle scale independent. Two possible choices for referent magnetic energy could be  $U_{\uparrow\uparrow}=10^{-18}$J and 67$\cdot10^{-18}$J, i.e., $256k_{\rm B}T$ and $1.6\cdot10^4k_{\rm B}T$, for particles with magnetic moments $m =$1.15 A\textmu m$^2$ and 9.2A\textmu m$^2$, respectively, where $T=300$K is the temperature and $k_{\rm B}$ Boltzmann's constant.\vspace{-1mm}}
  \label{fig:enAAAB}
  \end{figure}

  \begin{figure}[th!]
  \centering
  \includegraphics[width = 8 cm]{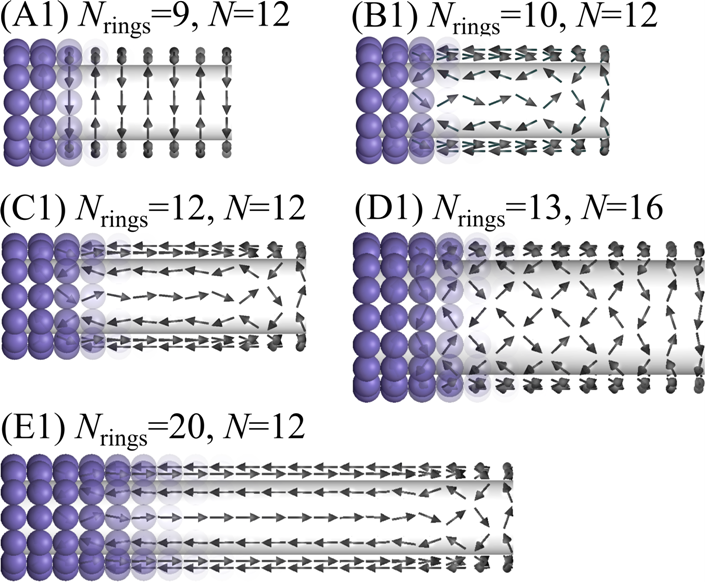}
  \caption{Illustrative examples of characteristic ground state magnetization for tubes with AA stacking.
  Configurations (A1), (B1), (C1) and (E1) are obtained with curvature $N=12$, and (D1) with $N=16$.\vspace{-1mm}}
  \label{fig:AAtube}
  \end{figure}

  \begin{figure}[th!]
  \centering
  \includegraphics[width = 9 cm]{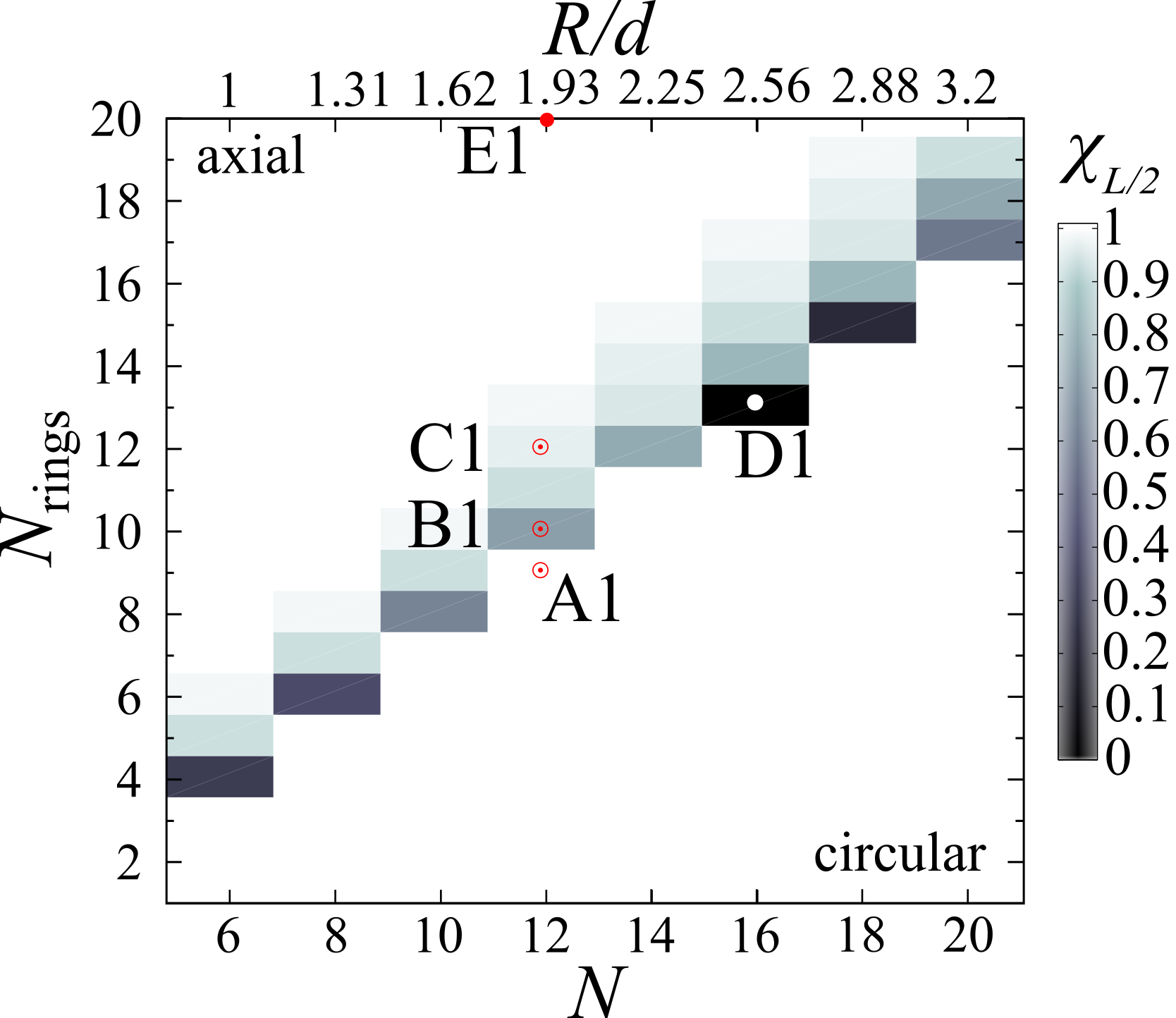}
  \caption{State diagram of AA tubes.
  It is shown in 2D tube length-curvature parameter space, i.e., $L\left(R\right)$  or $N_{\rm rings}\left(N\right)$,
  with clear indication of axial, circular and transitional vortex magnetization states.
  Coloring method based on order parameter $\chi_{L/2}$, defined in Eq.~\ref{chi}, is applied. The order parameter $\chi_{L/2}$ is zero in axial and circular magnetic states, i.e., when magnetic texture is parallel with tubes geometry, and equal to unity when magnetic structure is turned $45^\circ$ (i.e. equidistant from axial and circular magnetization).} 
  Configurations (A1), (B1), (C1), (D1), and (E1) are shown in Fig.~\ref{fig:AAtube}. The results are in principle scale independent. In this work, we have used in all examples length scale $d=$50\textmu m.\vspace{-1mm}
  \label{fig:AAphd}
  \end{figure}

  \begin{figure}[th!]
  \centering
  \includegraphics[width = 9cm]{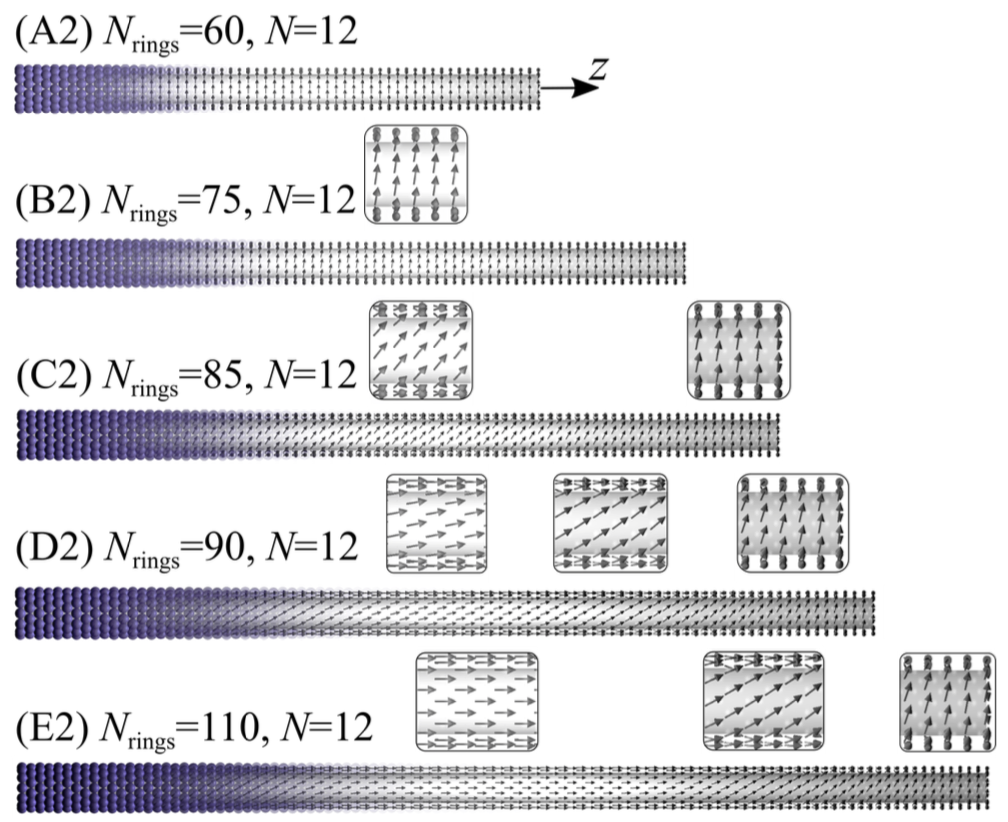}
  \caption{Illustrative examples of characteristic ground state magnetization for tubes with AB stacking.\vspace{-5mm}}
  \label{fig:ABtube}
  \end{figure}

  \begin{figure}[th!]
  \centering
  \includegraphics[width = 9cm]{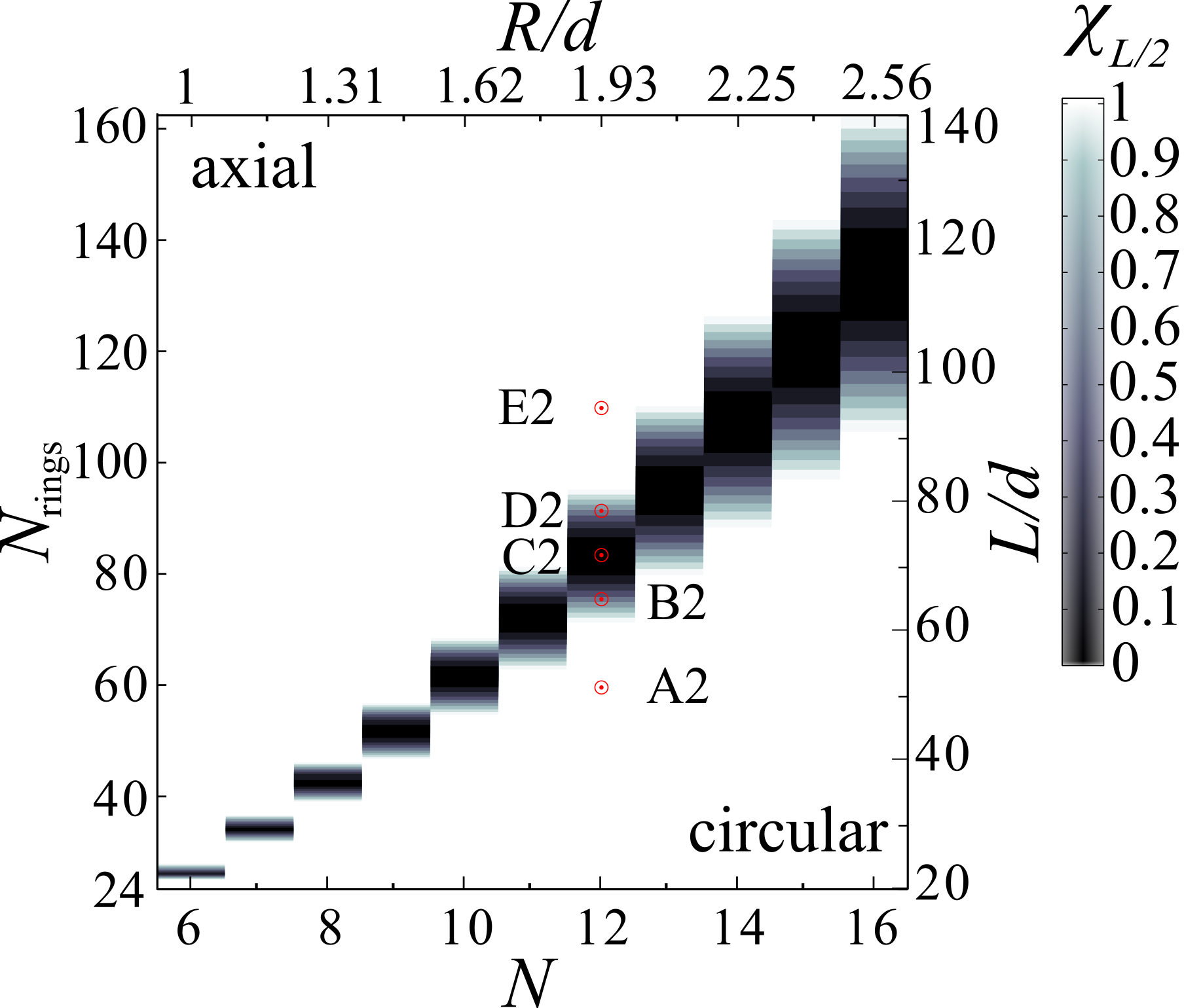}
  \caption{State diagram of AB tubes. It is shown in 2D tube length-curvature parameter space, i.e.,
  $L\left(R\right)$  or $N_{\rm rings}\left(N\right)$, with clear indication of axial, circular and transition helical states.
  Coloring method $\chi_{L/2}$, defined in Eq.~\ref{chi}, is applied. The order parameter $\chi_{L/2}$ is zero in axial and circular magnetic states, i.e., when magnetic texture is parallel with tubes geometry, and equal to unity when magnetic structure is turned $45^\circ$ (i.e. equidistant from axial and circular magnetization).} Points (A2), (B2), (C2), (D2) and (E2) from the state diagram  are chosen as
  illustrative examples in Fig.~\ref{fig:ABtube}. The results are in principle scale independent. We used length scale $d=$50\textmu m.\vspace{-1mm}
  \label{fig:ABphd}
  \end{figure}

\begin{figure}[th!]
\begin{center}
\includegraphics[width=8cm]{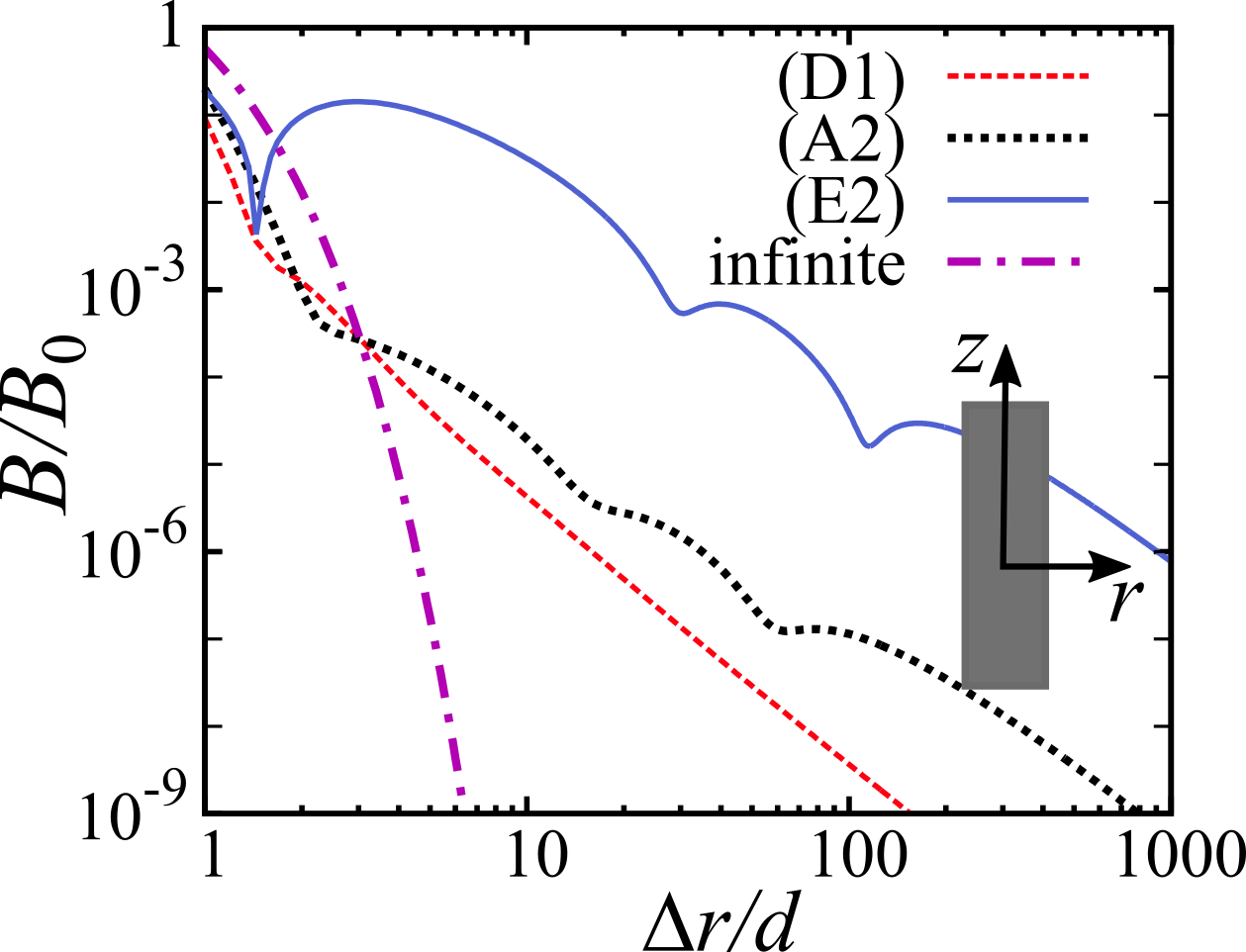}
\end{center}
\caption{Dependence of intensity of the magnetic field $B$ from radial distance $\Delta r=r-R$ from center of dipolar tubes, where $R$ is tubes radius. The magnetic field is given for (D1) AA tube in vortex state $L/d=13$ long and with curvature $N=16$, (A2) circularly magnetized AB tube with $L/d\approx53$ and $N=12$, (E2) axially magnetized AB tube with $L/d\approx98$ and $N=12$, and uniformly axially magnetized infinite AB tube with the same curvature ($N=12$). The distance $\Delta r/d=1$ is the distance of closest approach of the particle to the tube. The configurations of AA and AB tubes are shown in Figs.~\ref{fig:AAtube} and~\ref{fig:ABtube}, respectively. We used length scale $d=$50\textmu m. Referent magnetic field is $B_{\rm 0}=1$\textmu T for two particles with magnetic moments 1.15A\textmu m.} \label{fig:BB}
\end{figure}

\subsection{Magnetization states in finite tubes}
\label{phasediagram}

We will go one step further and consider finite tubes which consist of $N_{\rm rings}$ stacked rings. Tube's length influences ground state dipole orientation on both global and local levels. The competition between the two geometrical parameters, ($i$) curvature $N$ and ($ii$) tube length $N_{\rm rings}$, leads to different possible magnetic states of the tube. The energies of ground states at prescribed number of rings $N_{\rm rings}$ and for two curvatures $N=8, 12$ are given in Fig.~\ref{fig:enAAAB} for AA and AB tubes (i.e. square and triangular tubular structures).

\subparagraph*{Finite AA tubes} For square AA stacked tubes with $N=12$ curvature, circular magnetization state is stable for ($2 \leq N_{\rm rings} \leq 9$) rings. It turns out that circular magnetization is the ground state of short tubes, relative to their constitutive ring size $N$. The circular magnetization case $N_{\rm rings}=9$, is illustrated in Fig.~\ref{fig:AAtube}(A1). The change of magnetization towards axial is abrupt for $N_{\rm rings}=10$ and curvature $N=12$, see Fig. \ref{fig:AAtube}(B1). We observe a local antiferromagnetic circulation formed almost over the whole length except in terminal rings.
The dipoles in middle of the tube are only slightly misaligned with tube's axis (i.e., for angle $0.12\pi$). As result of change of magnetic order we observe, Fig.~\ref{fig:enAAAB}, that slope of cohesive energy changes from $N_{\rm rings}=9$ to $10$, i.e., between points (A1) $u_{\rm AA}^{12,9}=-2.4534$ and (B1) $u_{\rm AA}^{12,10}=-2.4589$. Extending further the tube length $N_{\rm rings}\geq13$, we observe well formed axial antiferromagnetic state with chains of alternating magnetization parallel to the tube axis.

The state diagram of AA tubes are given in Figure~\ref{fig:AAphd}. The calculated equilibrium states are given for different curvatures and lengths of AA.
Coloring method in state diagram is based on the local order parameter, conveniently defined as:
  \begin{equation}
  \chi_{L/2} = |2\langle (m_z/m)^2 \rangle_{L/2} - 1|,\label{chi}
  \end{equation}
where $(m_z/m)^2$ is scaled intensity of local magnetization in axial direction and $\langle\rangle_{L/2}$ is the average in the middle of the tube ($z = L/2$)~\footnote[8]{In case of even number of rings, i.e., $N_{\rm tot}=2k$, we take two rings above/below $z=L/2$.}. The idea of the order parameter is to visualize transition states (between axial and circular). Magnetic states which do not match with axial nor circular state in the middle of the tube are referred to also as {\it vortex states} in AA tubes.
The order parameter measures misalignment of $\vec{m}$ from the geometry of the tube, it is $\chi_{L/2} = 1$ in circular, i.e., $(m_z/m)^2=0$, and axial states, i.e., $(m_z/m)^2=1$, i.e., white areas comprising points (A1) and (E1) in Fig.~\ref{fig:AAphd}
(cf. also Fig.~\ref{fig:AAtube}). The state diagram contains three regions corresponding to the three classes of equilibrium states. We observe pure circular magnetization with no axial dipole component, for short tubes. 
In the transition state there is a change from the dominantly axial orientation of dipoles in the middle of the tube ($z = L/2$) to a vortex-like orientation at tube's ends ($z = {0, L}$).
We observe that a transition from a vortex to a axial state follows roughly linear trend for $4\le N\le14$.
For $N=16$ this trend is broken and the transition occurs earlier (after a single additional ring and not two). The resulting local order parameter is very small , $\chi_{L/2}\approx0$. This is all result of a strong local
circulation, i.e., $\theta=\pi/4, m_{\phi}=m_{\rm z}=m/\sqrt{2}$, cf. value of $\chi_{L/2}$ at point (D1) in Fig.~\ref{fig:AAphd} and also visualization in Fig.~\ref{fig:AAtube}.

\subparagraph*{Finite AB tubes} In the case of AB stacked tubes (triangular lattice), for $N=12$ curvature, circular state is stable for ($2 \leq N_{\rm rings} \lesssim 70$) rings. After that only dipoles in the middle of the tube start significantly
to change magnetization, cf. Fig.~\ref{fig:ABtube}(A2) and (B2). Only when local order parameter, $\chi_{L/2}\approx0$, we observe a change in dependence of cohesive energy on tube's length $N_{\rm rings}$,
cf. Fig.~\ref{fig:ABtube}(C2) and Fig.~\ref{fig:enAAAB} for $N_{\rm rings}=85$. The energy for configuration (C2) is $u_{\rm AB}^{12,85}=-2.6895$. The similarity of observed state transitions with increasing length
of the dipolar tube to state transition observed in solid magnetic nanotubes is striking ~\cite{LanderosPRB09,WyssARXIV17}. This is surprising due to the absence of the exchange interaction in dipolar tubes. 
The transition state $\chi_{L/2}\approx0$ we call {\it helical state}. The helical state, both in solid and dipolar tubes, is a result of the interplay between tube's curvature and length. We find three equivalent states: clock-wise, anti-clockwise and symmetric, within numerical accuracy, as a result of broken symmetry.

There are three clear differences between transitions from circular to axial state in AA and AB tubes, as seen in Figs.~\ref{fig:AAphd} and~\ref{fig:ABphd}:
\begin{itemize}
\item The transition occurs at smaller tube lengths in case of AA tubes, i.e., in AA-tubes for curvature $N=12$ transition is at $N_{\rm rings}\approx 10$ while for AB-tubes it will occur at at $N_{\rm rings}\approx 80$;
\item For AA tubes, circular state sharply changes into transitional vortex state when threshold length is reached. In the case of AB tubes, the transition through helical state is gradual with increasing length;
\item Edge effects at tube's ends, i.e., in vicinity of $z = (0, L)$, are much stronger in AB tubes than in the case of AA tubes, i.e., in AB-tubes for curvature $N=12$ they extend over $\Delta N_{\rm rings}=30$ rings on each side of the tube, compared to only up to $\Delta N_{\rm rings}=3$ rings.
\end{itemize}

It is insightful to compare
energies of obtained finite tubular magnetizations with the limits of an infinite planar triangular and square lattice. In the case of AB tubes for $N=12, N_{\rm rings}=200$, we obtain $u_{\rm AB}^{12,200}=-2.7203$ and
an energy deviation of about $15 \%$ from the infinite triangular plane case. This is essentially due to edge effects that are still non negligible. For much shorter AA tubes, i.e., $N=12, N_{\rm rings}=35$, we are with
$u_{\rm AB}^{12,35}=-2.5233$ within $10 \%$ from the infinite plane case.

At this point we would like to draw a comparison with solid-state MNTs. In MNTs magnetic properties are mainly defined by dipole-dipole and exchange interactions, wherein the later stems from quantum mechanical considerations. Exchange is a short range interaction that, in the micromagnetic approximation, is typically characterized by the exchange length ($l_{\text{ex}}$) that is not larger than a few tens of nanometers. The quantum mechanic signature in magnetic states of nanotubes can be neglected whether by choosing curvature $R\gg l_{\text{ex}}$ or reducing exchange length to zero. The magnetic equilibrium states of MNTs are mostly defined according to the ratio between MNT dimensions, such as its length $L$ and radius $R$. The radii $R_{\text{F}}\sim \eta\ l_{\text{ex}}$ and $R_{\text{V}}\approx \gamma\ l_{\text{ex}}$ are critical transition radii with $\eta=1-10$ and $\gamma=10-50$.  In MNTs with $L\geq R$ and $R< R_{\text{F}}$ uniform axial states are the preferred ground states. At $R_{\text{F}}<R<R_{\text{V}}$ and $L\gg R$ the magnetization is in axial state (i.e., only the center of the tube is axially magnetized), and if the length is reduced to $L\approx R$ magnetization turns into the circular state. The helical state appears as a transition state between the axial and circular state as a result of reduction of the tube's length. 
All these states have been predicted theoretically\cite{EscrigJMMM07,LanderosPRB09} and measured experimentally just recently\cite{WeberNL12,RufferNANOSCALE12,BuchterPRL13,BuchterPRB15,WyssARXIV17}. Thus, solid state MNTs with weak or comparable exchange interaction regarding the dipolar interaction) will exhibit circular magnetic order. This is not the case in dipolar tubes, consisting of discrete (nano- or even micro-particles), where exchange interaction is not present. And still, we could find all states seen in MNTs (circular, helical, and axial). We also observe similar tendencies with respect to the tube's size. We find the circular state in short, helical intermediary state for medium, and the axial state in long dipolar tubes.

The principal difference between the AA- and AB-tubes is the total magnetic moment. For AA-tubes total magnetic moment is zero. In the case of AB-tubes, the axial and helical states have a finite total magnetic moment, just like MNT counterparts. Figure~\ref{fig:BB} shows the dependence of magnetic field intensity on radial distance from the center of AA and AB tubes. The magnetic field at the closest approach of the probe particle $\Delta r/d=1$ is always smaller than the magnetic field of a single constitutive particle $B/B_0=1$ in side by side $\uparrow\downarrow$ configuration. The vortex state in AA tube results in a $B_{D1}(1)/B_0=0.09$, while in case AB tubes $B_{E2}(1)/B_0=0.19$. Smallest structure in Figure~\ref{fig:BB} has more than 200 constitutive particles, i.e., (D1) AA tube in the vortex state. In all three cases of the finite tubes, the intensity of magnetic field far from the tube follows power law on distance, i.e., $B\sim\Delta r^{-3}$ for $\Delta r/L>>1$. Only in axially magnetized infinite AB tubes magnetic field exponentially decays with distance $\Delta r/d$ and therefore fulfills flux closure, cf. also Ref.~\cite{TaoChainsInfinite}. This result is not surprising from the micromagnetic point of view since in a finite object a singularity-free solution cannot exist for topological
reasons\cite{feldtkeller2017continuous}. Only if the system is infinite at least in one dimension, micromagnetic solutions may be constructed. As result magnetostatic energy is minimized, leading to similar ground states in finite MNTs and dipolar tubes, that tend to reduces the stray field but cannot make it negligible.

\begin{figure}[h!]
\begin{center}
\includegraphics[width=8cm]{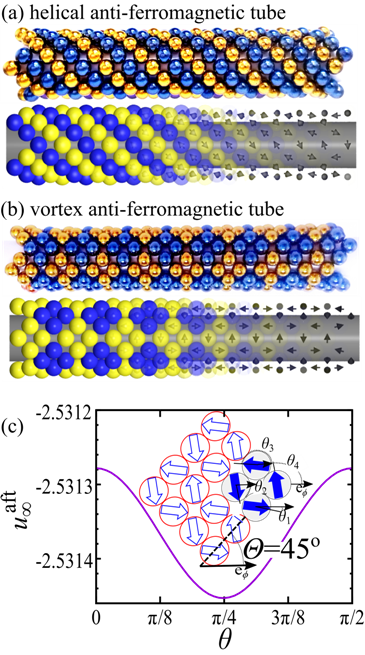}
\end{center}
\caption{The helical (a) and vortex (b) anti-ferromagnetic states realized with neodymium magnets (in upper panels) and magnetization pattern. Energy spectrum of configurations (c) for same infinite configuration with respect to angle $\theta$, where $\theta_1=\theta$, $\theta_2=-\theta$, $\theta_3=\pi+\theta$, and $\theta_4=\pi-\theta$. Case $\theta=\pi/4$ corresponds to helical and $\theta=0, \pi/2$ vortex state. The chiral angle, i.e., the angle between tread of particles and tangent to cylinder radius, is $\Theta=45^\circ$ also marked in the planar scheme of the system in panel (c). To convert results in real units, for example the referent magnetic energies $U_{\uparrow\uparrow}=10^{-18}$J or 67$\cdot10^{-18}$J could be used that correspond to $256k_{\rm B}T$ or $1.6\cdot10^4k_{\rm B}T$, in cases of particles with magnetic moments $m =$1.15 A\textmu m$^2$ or 9.2A\textmu m$^2$, respectively, where $T=300$K is the temperature and $k_{\rm B}$ Boltzmann's constant.} \label{fig:2}
\end{figure}

\subsection{Chirality and degeneracy breakup}

In this section, we would like to point out, how chirality of the structure influences energy barriers between different states in dipolar tubes. The ribbons of assembled particles can be rolled at different ("chiral") angles $\Theta$. In our self assembly experiment, combination of magnetic field along the wire and circular electromagnetic field will result in creating ferromagnetic tubes with specific chiral angle. We will only briefly analyse only limiting cases which are actually most interesting ones from the point of the metastability (i.e., energy differences between different states). Antiferromagnetic tubes need to be created on a tubular structure with tilted pattern. The radius of these tubes will depend on the lattice structure (i.e., square or triangular) and chiral angle. Here we will demonstrate how combination of the rolling angle and radius decides the tube's energy. In previous section, we have calculated energies of different states for tubes obtained by stacking of the rings. In the following text, we will follow energy gains and losses due to change of chirality (orientation) of tube's lattice with its axis $\Theta=30^\circ$ in triangular and $\Theta=45^\circ$ in square lattice. 

First, we will compare the energy of infinite AB tube in Fig.~\ref{fig:enAng}(b) and ZZ tube shown in Fig.~\ref{fig:addtubes}(c). While ZZ-tube is aligned with the tube's axis, ribbon generating AB tube is rolled under 60$^\circ$ angle, see  Fig.~\ref{fig:addtubes}(b). The energy of circular state in Fig.~\ref{fig:enAng}(b), $u_{\rm AB}^{\rm circular}=-2.694$, for unit ring of $N=12$ particles and $\theta=0$. Helical state ($\theta=\pi/3$), in Fig.~\ref{fig:enAng}(a), is more energetically favorable, $u_{\rm AB}^{\rm helical}=-2.7315$. Axially magnetized state has energy $u_{\rm AB}^{\rm axial}=-2.7441$ for $\theta=\pi/2$. Difference between circular and axial state energies is small, i.e., less than $2\%$ of total energy.  Already at moderate curvatures, i.e., $R/d=1.932$, the difference to infinite triangular plane value ($u_{\rm AB}^{\infty}$) is small, $u_{\rm AB}^{\rm axial}-u_{\rm AB}^{\infty}\approx0.015$ or roughly $0.5\%$ of total energy value. If we chose chirality to align tube's structure with its axis, as in ZZ tubes shown in Fig.~\ref{fig:addtubes}(c)\footnote{In ZZ tube, particles form chains (so called, filaments) parallel to the tube's axis.}, the energy converges faster to infinite triangular plane value. The energy difference, for system in Fig.~\ref{fig:addtubes}(c), is  $u_{\rm ZZ}^{\rm axial}-u_{\rm AB}^{\infty}\approx0.001$. Improved convergence of axial state comes with a marginal increase of energy difference to circular and helical states of less than $3\%$ of total energy value. The energies of circular and helical states, for the ZZtube in Fig.~\ref{fig:addtubes}(c), are $u_{\rm 2}^{\rm circular}=-2.618$ and $u_{\rm 2}^{\rm helical}=-2.7$, respectively. We can conclude that by changing chirality we can manipulate energy differences between different states. 

The AA tube's square lattice is aligned with tube's axis, see Fig.~\ref{fig:enAng}(a). What will it happen if we turn the tube's lattice structure $45^{\circ}$? We show the configurations and results of energy calculations in Figure~\ref{fig:2}. To demonstrate stability of the structure it is realized also with neodymium magnetic spheres. The striking feature is a comparably small energy increase of $u_{\infty}^{\rm helical}-u_{\infty}^{\rm vortex}=1.8~ 10^{-4}$, cf. Figure~\ref{fig:2}(c). This means that in realization with a finite temperature this system would be degenerate. Since  an infinite tube can never be realized one could ask how significant is the influence of the edges? In this context we calculate energies of finite tubes consisting of $N=208$ particles, i.e., which correspond exactly to the helical and vortex configurations shown in Figure~\ref{fig:2}(A) and (B), and obtain values $u^{\rm vortex}_{208}=-2.4527$ and $u^{\rm helical}_{208}=-2.4495$, respectively. Therefore at least in these two finite configurations the energy is relatively close to each other (within $2\%$) and to that of infinite tubes (i.e., within $5\%$). In contrast to AA tubes in previous section, local magnetic order of finite and infinite tube's in Figure~\ref{fig:2} is quite similar. The reason is that tubes finish crown (zig-zag) ring which prevents formation of a continuous head-tail magnetic order. 

\section{Conclusion and outlook}
\label{conclusion}

In the first part of the paper, we demonstrate that using magnetic particles with permanent dipolar moment gives additional design freedom for an experiment recently proposed. Injection of an electrical current in a conductor wire induces a electromagnetic field. The radial gradient of this field owns the ability to attract magnetic beads. The particles, therefore, assemble on the wire surface. We explored the intensity of the electromagnetic field that leads to a transformation of the clusters attached to the wire into a single layer tubular structure. We further analyse the limits on the injected currents to minimize the Joule heating and steer the particle assembly. In this regard, we have found a realistic range of current and resulting electromagnetic field at which the assembly of spheres and its magnetic orientation are stable and controllable. Our results are generic and can be scaled to many different systems. Once the current is switched off, the circular electromagnetic field disappears, and the particles stay assembled held by interparticle interactions. From this point on, the magnetization of colloidal particles turns and relaxes to the equilibrium configuration.

In the second part of the paper, we studied the curvature-induced breakup of the continuously degenerated state when a two-dimensional ribbon of spheres is curved and transferred to the cylinder. We show that different ferromagnetic states, observed previously~\cite{Messina_PhysRevE_2014,StankovicSM2016} are result of curvature induced energy barriers that lift the continuous degeneracy in the triangular lattice. We performed a systematic investigation of the degeneracy break-up as a function of the tube length and packing symmetry (square or triangular), which lead to a number of equilibrium magnetic states of dipolar tubes. For triangular packing, we show that dipolar tubes transcend scale. Its equilibrium states mimic the ground magnetization of magnetic nanotubes where the dipolar interaction is whether comparable or dominate over the exchange interaction. Indeed, we found the circular state in short tubes, the axial state in long tubes, and the helical state in between. This is important conclusion since it shows that all these states could exist in magnetic nanotubes also without  exchange interactions. We find that planar square lattice has a continuously degenerate antiferromagnetic state. In tubes with square lattice, we have found remarkable magnetic vortex configurations formed spontaneously. Such configuration was observed previously only in system of magnetic cubes due to intricate relation between crystalline of the cubes and packing. Antiferromagnetic states have no analogous in the set of magnetic ground states in continuum magnetic nanotubes~\cite{LanderosPRB09,StreubelJPDAP16} and are remarkable due to their curvature-induced stability and non-colinear texture. Indeed, these non-colinear states can be very attractive for further research about magnetization dynamics (reversal processes mediated by domain wall propagation and spin-waves) due to the macroscopic scale of dipolar tubes, therefore less complexity in experiments.   

In the context of curvilinear nanomagnetism,\citep{LanderosPRB09,OtaloraPRL16} the present theoretical result could represent a departure point and alternative mean to test and explore equilibrium and dynamic magnetic properties at macroscopic scales. The dipolar tubes present an alternative technique to reduce the complexity of experiments and a platform to prove concepts for applications in magnonics at more accessible parameters (reduced frequencies and macroscopic wavelengths). The experimental realization of cylindrical magnetic objects is very demanding since curvature effects might be overshadowed by wall pinning on imperfections, such as grain boundaries and edges~\cite{StreubelJPDAP16}. The realization of presented quasi-one-dimensional or edge free ferromagnets and antiferromagnets would create an accessible platform for testing concepts of spin-based electronics (e.g., Cherenkov-like spin wave emission~\cite{YanAPL11,OtaloraAPL12} or curvature induced non-reciprocities in magnonics~\cite{OtaloraPRL16,OtaloraPRB18}) and information technologies by getting around a requirement of robust magnetic uniformity at temperatures of technological relevance. An additional application of the tubular assemblies of dipoles could be modeling of the ordered planar systems due to the absence of the lateral edges in curved geometry and low energy barriers. Macroscopic dipoles were, for instance, successfully used to study frustrated states in spin glasses~\cite{MelladoPRL12}. Although the previous approach can not be applied straightforward to ordered planar structures, since edge effects could overshadow properties like the response to magnetization reversal, it can motivate further studies on this topic.


\section{Acknowledgements}

The authors acknowledge financial support from European Commission H2020 project DAFNEOX (Grant No. 645658).
Numerical calculations were run on the PARADOX supercomputing facility at the Scientific Computing Laboratory of the Institute of Physics Belgrade.
I.S. and M.D. acknowledge support of Ministry of Education, Science, and Technological Development of Republic of Serbia - projects ON171017 and III45018. I.S. and C.G. acknowledge the financial support received by Proyecto CONICYT PIA/Basal FB 0821 and CONICYT MEC80170122.

\bibliography{arxiv} 

\end{document}